\documentclass[useAMS,usenatbib,aps,prd,twocolumn]{revtex4-1}
\usepackage{times,graphicx,amsmath,amsfonts,amssymb,aas_macros,epstopdf,ulem}
\usepackage{epsfig}
\usepackage[usenames,dvipsnames]{xcolor}
\usepackage{url}
\usepackage{subfigure}
\usepackage[colorlinks=true, urlcolor=black, citecolor=blue]{hyperref}
\usepackage[utf8]{inputenc}

\def\ie{{\frenchspacing\it i.e.}}
\def\eg{{\frenchspacing\it e.g.}}

\def\be{\begin{equation}}
\def\ee{\end{equation}}
\def\ba{\begin{eqnarray}}
\def\ea{\end{eqnarray}} 
\def\Msun{h^{-1}{\rm M}_{\odot}}
\def\hmpc{h^{-1}\,{\rm Mpc}}
\def\mpch{h\,{\rm Mpc}^{-1}}
\def\kms{\, {\rm km }\, {\rm s}^{-1}}
\def\hmpcc{h^{-3}\,{\rm Mpc}^3}
\def\hkpc{h^{-1}\,{\rm kpc}}
\def\dd{\textrm{d}}
\def\lcdm{\Lambda{\rm CDM}}
\def\de{\delta}

\def\ln{{\rm ln}\,}

\def\frac#1#2{{\textstyle{#1\over #2}}}

\def\simlt{\stackrel{<}{{}_\sim}}
\def\simgt{\stackrel{>}{{}_\sim}}

\def\BF{\ensuremath{\mathcal{B}}}
\def\VD{\ensuremath{\mathcal{D}}}
\def\MN{\ensuremath{\mathcal{M}}}
\def\lcdm{\ensuremath{\Lambda}CDM}

\def\sam_main{\ensuremath{\langle n\rangle=6\times10^{-3}h^3}Mpc\ensuremath{^{-3}}}
\def\samI{\ensuremath{\langle n\rangle=5\times10^{-4}h^3}Mpc\ensuremath{^{-3}}}
\def\samII{\ensuremath{\langle n\rangle=5\times10^{-5}h^3}Mpc\ensuremath{^{-3}}}
\def\samIII{\ensuremath{\langle n\rangle=5\times10^{-6}h^3}Mpc\ensuremath{^{-3}}}

\newcommand{\LGO}{\textsc{\bf LGO}}
\newcommand{\RNDO}{\textsc{\bf RNDO}}

\begin{document}
\title{Uneven flows: On cosmic bulk flows, local observers, and gravity}
\author{Wojciech A.~Hellwing$^{1}$}
\author{Maciej Bilicki$^{2,3}$}
\author{Noam I.~Libeskind$^{4}$}
\affiliation{$^{1}$Center for Theoretical Physics, Polish Academy of Sciences, Al. Lotników 32/46, 02-668 Warsaw, Poland}
\affiliation{$^{2}$Leiden Observatory, Leiden University, Niels Bohrweg 2, NL-2333 CA Leiden, The Netherlands}
\affiliation{$^{3}$National Centre for Nuclear Research, Astrophysics Division, P.O.Box 447, 90-950 \L{}\'{o}d\'{z}, Poland}
\affiliation{$^{4}$Leibniz-Institut f\"ur Astrophysik Potsdam (AIP), An der Sternwarte 16, D-14482 Potsdam, Germany}
\date{\today}
\begin{abstract}
Using N-body simulations we study the impact of various systematic effects on the low-order moments of the cosmic velocity field: the bulk flow (BF)
and the Cosmic Mach Number (CMN). We consider two types of systematics: those related to survey properties and those induced by observer's
location in the Universe. In the former category we model sparse sampling, velocity errors, and survey incompleteness 
(radial and geometrical).
In the latter, we consider Local Group (LG) analogue observers, placed in a specific location within the Cosmic Web, satisfying
various observational criteria.
We differentiate such 
LG observers from Copernican ones, who are at random locations. We report strong systematic effects on the measured BF and CMN induced by 
sparse sampling, velocity errors and radial incompleteness. For BF most of these effects exceed 10\% 
for scales $R\simlt 100\hmpc$. For CMN some of these systematics can be catastrophically large (\ie{} $>50\%$) also on bigger scales. 
Moreover, we find that the position of the observer {in the Cosmic Web} significantly affects the locally measured BF (CMN), 
with effects as large as $\sim20\%$ ($30\%)$
at $R\simlt50\hmpc$ for a LG-like observer as compared to a random one. {This effect is comparable to the sample variance at the same scales.}
Such location-dependent effects have not been considered previously in BF and CMN studies and here we report their magnitude and scale
for the first time. To highlight the importance of these systematics,
we additionally study a model of modified gravity with $\sim15\%$ enhanced growth rate (compared to general relativity).
We found that the systematic effects can mimic the modified gravity signal. The worst-case scenario is realized for a case of a 
LG-like observer, when the effects induced by local structures are degenerate with the enhanced growth rate fostered by modified gravity.
Our results indicate that dedicated constrained simulations and realistic mock galaxy catalogs will be absolutely necessary to fully benefit from
the statistical power of the forthcoming peculiar velocity data from surveys such as TAIPAN, WALLABY, Cosmic Flows-4 and SKA.
\end{abstract}

\maketitle

\section{Introduction}
\label{sec:intro}
The standard model of cosmology -- Lambda Cold Dark Matter (LCDM) - is extremely successful in explaining a plethora of observations. 
These include the features of the Cosmic Microwave Background \citep[\eg][]{WMAP9,Planck15}, the primordial nucleosynthesis and light 
element abundances \citep{Yang1984,Walker1991}, 
the growth of primordial density perturbations into the present-day large-scale structure (LSS) \citep{Percival2001,Tegmark2004,Beutler2017}, 
as well as 
the late-time accelerated expansion of the Universe \citep{acceleration1,acceleration2,Percival2010,Weinberg2013}. 
However, since LCDM is mostly phenomenological in its nature, this spectacular success comes at a price of accepting that the main contributors 
to the cosmic energy budget are dark matter (DM) and dark energy (DE), which have not been directly detected in any experiments so far. 
Therefore, it is desirable to look for other probes of the cosmological model, especially those which do not share at least some of
the systematics of the aforementioned measurements.

In this context, the peculiar motions of galaxies -- \ie\ deviations from the uniform Hubble flow -- are considered as particularly valuable 
\cite{Strauss1995,Koda2014,Howlett2017}. Induced by gravity only, they are not affected by such systematics as galaxy bias, 
which plagues for instance the measurements of galaxy clustering. Peculiar velocities can be therefore used, at least in principle, 
to obtain constraints on various cosmological parameters such as the mean matter density or the growth of structure \cite{Nusser2011a,Hudson2012}, 
independently of other methods.

Arguably the most popular statistics of the velocity field is the bulk flow (BF), \ie\ the net peculiar motion of galaxies contained in a given volume. 
BF probes large-scale fluctuations of matter distribution, and should generally diminish with increased volume. 
Over the decades, BF measurements have often been subject to various controversies. An example from early studies is by 
\cite{Lauer1994}, who measured a net motion of Abell clusters amounting to $\sim700 \kms$ within a radius of 15,000 $\kms$, 
which was however not confirmed by subsequent analyses \cite[\eg][]{Giovanelli1996,Dale1999} (but see \cite{Hudson1999}).
More recently, \cite{Watkins2009} claimed significant BF ($\sim 400 \kms$) on scales of $\sim100\hmpc$ from a combined sample 
of galaxies and clusters, which also is not supported by several other studies \cite[\eg][]{Nusser2011a,Hoffman2015} 
(see however \cite{Watkins2015}). Even more controversial are the claims of the very large scale ($\sim 300\hmpc$) `dark flow' 
by \cite{Kashlinsky2008}, which again is not corroborated by related analyses \cite{Osborne2011,Planck2014XIII}. 
Thanks to the ever growing {amount of} observational data, there is continued interest in measuring the BF and, if these discrepancies could 
be resolved, using it as a cosmological probe; 
for some more recent results see 
\cite{Colin2011,Nusser2011b,Weyant2011,Branchini2012,Mody2012,Turnbull2012,Feindt2013,Lavaux2013,Ma2013,
Rathaus2013,Feix2014,Hong2014,Ma2014,Appleby2015,Huterer2015,Scrimgeour2016,Springob2016}. 

Part of the BF `controversy' (or more precisely, inconsistency between some measurements) is due to the fact that 
many of the BF assessments are not directly comparable due to different estimators used, with specific sensitivity to various 
scales and systematics \cite{Nusser2014,Andersen2016,Nusser2016}. The quality and volume of the velocity data is another important issue here.
We note that some of the developed estimators do not use peculiar velocities at all to estimate the BF 
\cite[\eg][]{Kashlinsky2008,Nusser2011b,Branchini2012,Mody2012,Lavaux2013,Feix2014}, they are thus not sensitive to the related biases, 
although this of course does not make them immune to other, often major, systematic effects.

The BF continues to be regarded as a promising probe of cosmology especially taking into 
account that larger, denser, and more accurate samples of peculiar velocities are expected to appear in 
the coming years from such 
surveys as Taipan \cite{Taipan}, WALLABY \cite{WALLABY}, or CosmicFlows-4 \cite{CF-3}. However, agreement is gradually building 
up that in order to take full advantage of these future datasets for BF and other velocity-based measurements, the control of systematic
effects and biases is crucial for proper data interpretation. Recent developments of {\eg \cite{Andersen2016,Hellwing2017,LiPan2012}} highlight
the importance of selection and observer-driven effects {for peculiar velocity studies. Ref.} 
\cite{Andersen2016} considered the impact of purely geometrical selection effects
on the inferred bulk flows, including the partial sky coverage. In addition, Ref. \cite{LiPan2012} investigated mainly the effect of
different radial selection functions and the corresponding galaxy/halo weighting. Both works report 
{the} importance of these two systematic
effects that can bias the data, but the effect they {studied referred to} a hypothetical Copernican observer. 
The results of \cite{Hellwing2017} however underline that for relatively shallow and sparse velocity data, the specific location
of the observer within the cosmic web affects in a non-trivial way the cosmic variance of the velocity observables.
Inspired by these previous results,
in this work we will readdress this issue by looking closely at the impact 
of the observer location (\ie\ importance of the local cosmic structures) on the inferred BF and related statistics.
We will show that the BF itself is very sensitive to such effects, which must be therefore properly accounted for 
when measuring it from the current and forthcoming datasets.

A statistics related to the BF, which uses additionally the third moment of the peculiar velocity field, is the cosmic 
Mach number (CMN) defined as the ratio of the BF to the peculiar velocity dispersion in the same volume \cite{Ostriker1990}. 
In the original proposal, CMN was regarded as a `critical test for current theories', and more recently quoted as `a sensitive 
probe for the growth of structure' \cite{Ma2012}. For other theoretical and observational studies of CMN and its importance 
for cosmology, see \cite{Suto1992,Strauss1993,Nagamine2001,AtrioBarandela2004,Agarwal2013}. In this paper we will examine 
the sensitivity of the CMN to the same systematics as those studies for the BF. Similar conclusions regarding the 
importance of such effects for CMN as in the case of BF will apply.

The cosmic velocity field reflects a continuous action of gravity integrated over the history of large-scale-structure growth. 
Thus it offers, in principle, a very sensitive probe of the very nature of gravity itself. Here, even small possible deviations from 
a general relativity (GR)-like force law provide minute galaxy
acceleration changes that are amplified when integrated over time. This has been shown by other authors for a range of velocity 
field statistics and viable modified gravity (MG) models (see \eg{}\cite{LiHell2013,Hellwing2014PhRvL,Ivarsen2016}). 
Thus, if one is able to control various systematic effects,
and in the case of known (assumed) cosmological parameters like $\Omega_m$ and $\sigma_8$ (taken for example from CMB observations), then
the galaxy velocity field (and its low-order moments) provides a potentially powerful way of 
constraining non-GR models. Such constraints
would foster an independent, thus complimentary, way of testing GR and measuring the local value of the growth rate, $f\equiv \ln D_+/\ln a$
\cite{Koda2014,Howlett2017}. {In order to be able to use the velocity data for testing gravity one needs to recognize and control
all important systematic effects. Consequently, in this paper we also consider a modified gravity model 
(deviating by $\sim15\%$ in the growth rate from GR) and compare its signal with the magnitude of various systematics in the GR case.}

{As briefly indicated above, various systematic and statistical effects that disturb the velocity data were a subject of careful 
study in the past.
However, except for the early work of Ref.~\cite{Tormen1993}, analyses of the impact of a specific location of the observer 
within the large-scale structure were not conducted. Ref.~\cite{Tormen1993} studied only the 2-point velocity statistics and they 
did not require the presence 
of any nearby Cosmic Web structures such as the Virgo cluster. Here, we will conduct a joint study of various systematic effects, starting from
sparse sampling and radial selection, up to the impact of the observer location. 
We will identify scales and magnitudes of various effects and compare them against expected statistical fluctuations in a systematic fashion. 
In this way we will obtain insight into scales, magnitudes and 
the interdependence of various systematical effects troubling BF and CMN measurements. This will constitute another important step for peculiar velocity studies towards the precision cosmology era.}

{The paper is organized as follows: in \S\ref{sec:simulations} we describe in details computer simulations used in this study; in \S\ref{sec:theory}
some theoretical preliminaries and relevant considerations are given; \S\ref{sec:mocks_non_linear} contains description of mock catalogs
and various observational effects that we model; in \S\ref{sec:obs-independent} we discuss the effects induced by systematics independent from
a specific observer's position, while in \S\ref{sec:LGO} the focus is given to signals measured by Local Group-analogue observers;
\S\ref{sec:gravity} compares signal from a modified gravity model with known GR systematic effects.
Finally \S\ref{sec:sumarry} summarizes our findings, this is followed by \S\ref{sec:discussion} where discussion and conclusions are given.
{Some additional tests and discussion about the influence of the simulation box size are given in the Appendix.}
}

\section{Simulations}
\label{sec:simulations}

To study cosmic flows we employ a set of large N-body simulations conducted with the use of the {\tt ECOSMOG} code \cite{ECOSMOG}. 
Time evolution of cosmic structures is here followed with respect to a background spatially flat Universe
described by cosmological parameters consistent with the 2013 results from the Planck mission \cite{Planck13}.
We imposed the following values: $\sigma_8=0.831$, $\Omega_m=0.315$, $\Omega_{\Lambda}=0.684$, $n_s=0.96$.
The growth of density fluctuations is modeled by assuming that all non-relativistic matter is collisionless, \ie\ we 
treat the baryonic component as DM. Ignoring baryonic physics will not introduce any significant biases as long as 
we are not interested in internal properties of individual halos but only in their spatial distribution and peculiar velocities
(\eg \cite{HellwingEAGLE2016}).Thus in our simulation
we place $1400^3$ DM particles in a cubic box of comoving size
of $1000\hmpc$. This particular set-up fixes the mass resolution at {$m_p=3.2\times10^{10}\Msun$. {\tt ECOSMOG}} is an extension
of the {\tt RAMSES} code \cite{RAMSES} and uses adaptive mesh refinement (AMR) and dynamical grid relaxation methods to compute
the gravitational potential and forces. Thus our simulations are not characterized by a single fixed force resolution, but to
gauge our dynamical spatial resolution we can use the cell size of the most refined AMR grid. In all our runs such a
grid had a rank of $N17$ resulting in the finest force resolution of $\varepsilon=7.6\hkpc$.

In this paper we aim to study various systematic effects that affect the lowest moments of the cosmic velocity field. 
For that reason we will be mostly concerned with {the} fiducial cosmological $\lcdm$ model. However, in Sec.\ref{sec:gravity} we will
compare the magnitude of various systematics with the predicted amplitude of a non-GR signature expected in the case
of a modified gravity model. As a representative guinea-pig we chose the so-called normal branch of Dvali-Gabadadze-Poratti 
(henceforth nDGP) model \cite{DGP2000,schmidt2009}. For a more detailed description of that model and its implementation in simulations, 
see the relevant Section.

Real astronomical observations measure the radial component of the peculiar velocity of a galaxy rather than of its host halo. Our simulations do not 
attempt to model assembly of galaxies in any way, but we can safely use the bulk velocities of DM halos found in our simulations
as faithful proxies for real galaxy peculiar velocities. This is the case since the studies of other authors {\citep[\eg][]{HellwingEAGLE2016,Ye2017}}
have shown that for central galaxies residing in massive halos their relative velocities with respect to their host halos are very small 
($\ie \leq 5 \kms$)
compared to the bulk-flow magnitude we will study here. In addition we do not expect that any non-zero galaxy velocity in relation to its host halo
would be correlated to the large-scale matter distribution which induces bulk flows.
Thus, due to global isotropy these velocities should average-out to zero
for scales much larger than a given halo radius.

To identify halos and subhalos in the simulations we employ {\tt ROCKSTAR} \cite{ROCKSTAR}, a phase-space friends-of-friends halo finder. 
To define a halo edge
we use the virial radius $R_{200}$, defined as the radius within which the enclosed density is $200\times \rho_c$, where $\rho_c$ is 
the critical cosmic density. For further analysis we keep all gravitationally bound halos that contain at least 20 DM particles each.
This sets our minimal halo mass to $M_{min}=20\times m_p=6.4\times10^{11}\Msun$. Based on the initial halo catalogs, we build 
our test halo populations
by distinguishing the central halos from satellites (subhalos).
For further analysis we keep only {the centrals}, which we will treat as rough mocks for population of central galaxies.
To obtain additional halo samples with lower number densities we {perform} random subsampling.
Our main catalog includes all central halos resolved in our simulation at $z=0$ and has a number density of
\sam_main. To obtain {sparser} samples we consequently dilute this main sample by randomly (and spatially uniformly) keeping only every $n$-th
halo. Thus we also obtain the following samples: \samI, \samII and \samIII.

Finally, since peculiar velocity catalogs are rather shallow, rarely reaching at present deeper than $\sim200\hmpc$ 
(see \eg\ \cite{Springob2007,Springob2014,Springob2016,CF-3}) we constrain all our analysis 
only to the $z=0$ snapshot of our simulations and to scales up to {$250\hmpc$. This being said, it is also imperative
to comment on the convergence of the simulation results at large scales. The velocity field is much more sensitive
to contributions from perturbation modes much larger than a given scale one considers. In other words, we can expect
that the finite-volume effects will be more pronounced here than in the case of the density field. To check what
scales we can trust, we have run additional tests involving three more simplified simulations with a varying box size.
The details and analysis of these are given in the Appendix. The results of these tests indicate
that on scales $R\simgt 200\hmpc$ the amplitude of our BF is systematically biased down by $15\%$
or more. However, the size of various systematic effects expressed as a relative BF magnitude difference appears
to be only weakly affected by the box size up to $R\sim250\hmpc$. This supports our choice
of the maximal scale we consider in this paper.}

\section{Theoretical preliminaries}
\label{sec:theory}

Throughout our work we assume the homogeneous and isotropic cosmological model, in which the background obeys 
Friedman-Lem\^aitre equations with a scale factor $a(t)$. All the quantities will be expressed in comoving coordinates \ie\ $\vec{x}={\vec{r} / a(t)}$. 
For background density $\rho_b(t)$ and density contrast $\de(\vec{x},t) \equiv {\rho(\vec{x},t) / \rho_b(t)-1}$, the Poisson equation linking 
the peculiar gravitational potential $\phi(\vec{x},t)$ with density perturbations is
\be
\label{eqn:Poisson-eqn}
\nabla^2\phi(\vec{x},t) = 4\pi G\rho_b(t)a^2\de(\vec{x},t)\,.
\ee
By integration, we obtain the expression for peculiar accelerations $\vec{g}$ \cite{1980Peebles}
\be
\label{eqn:potential-acceleration}
\vec{g}(\vec{x})=-{\nabla\phi\over a}=Ga\rho_0\int{\de(\vec{x^{'}})(\vec{x^{'}}-\vec{x})\over |\vec{x^{'}}-\vec{x}|^3} \textrm{d}\vec{x^{'}}\,.
\ee
Peculiar velocities $\vec{v}(\vec{x},t)$, defined as deviations from the Hubble flow, are coupled to the density field via the continuity equation:
\be
\label{eqn:cont}
{\partial \de\over\partial t}+{1\over a}\nabla\cdot[(1+\de)\vec{v}]=0\,.
\ee

\subsection{Linear theory predictions}
\label{subsec:lin_theory}

We can model the cosmic velocity field by performing a decomposition of the full three-dimensional (3D) field into a sum of longitudinal (non-rotational)
and transverse (rotational) components:
\ba
\label{eqn:velocity_decomp}
\vec{v}=\vec{v_L}+\vec{v_T}\,\,, \textrm{where:}\\
\nabla\times \vec{v_L}=0\,\,\textrm{and}\,\, \nabla\cdot\vec{v_T}=0\,.
\ea
In the linear regime, the velocity field is curl-free, thus $\vec{v_T}=0$ and the field is purely potential. Henceforth it can
be expressed as a gradient of a scalar function $\Psi_v$ (called the velocity potential)
\be
\label{eqn:vel_pot}
\vec{v}=-\nabla\Psi_v/a\,.
\ee
Now considering the continuity equation \eqref{eqn:cont} it can be shown that the velocity potential obeys
\be
\label{eqn:vel_pot_evo}
\nabla^2\Psi_v=Hfa^2\de\,\,,
\ee
where we have used the definition of the dimensionless growth rate $f\equiv \dd \log{D_1}/\dd \log{a}$. The growth rate only very weakly
depends on $\Lambda$ \cite{Lahav1991} and for a flat LCDM universe $f\approx \Omega_m^{0.55}$ \cite{Linder2005}. 
However, in general for some alternative
cosmologies (like coupled DE or modified gravity) it can take a different value and also be a scale-dependent function.

Finally, in the linear regime we have {$\phi\propto\Psi_v$} and $\vec{v}\propto\vec{g}$ 
(where $\vec{g}$ is the peculiar gravitational acceleration), and in particular {at $z=0$} one has 
\be
\label{eqn:linear_v_g}
\vec{v}={H_0 f\over 4\pi G\rho_0}\vec{g}={2f\over 3H_0\Omega_m}\vec{g}\,.
\ee
In the linear regime we can also express the relation between the power spectrum of density fluctuations, $P(k)\equiv\langle\de_k\de_k^*\rangle$,
and the dimensionless expansion scalar, $\theta_k$, which is the scaled velocity divergence (also called the expansion scalar)
\be
\label{eqn:expansion_scalar}
\theta\equiv {-\nabla\vec{v}\over aH_0}\,\,,\,\,\textrm{and}\,\,\vec{v}_k=aH_0\,{i\vec{k}\over k^2}\theta_k\,\,,
\textrm{so}\,\,P_{\theta\theta}(k)=\langle\theta_k\theta_k^*\rangle\,.
\ee
In the linear regime for a potential flow it follows from the continuity equation \eqref{eqn:cont} that
\be
\label{eqn:pk_ptt_equality}
P(k)=f^{-2}P_{\theta\theta}(k)\,.
\ee
The above relation is often used in the literature to approximate velocity power spectrum by
linear velocity divergence, thus neglecting dispersion and vorticity (see \eg \cite{Strauss1995}). Such approximation however
holds only on sufficiently large scales; those scales are generally larger {(\ie{} $\geq 60-100\hmpc$)} 
than in relevant analyses of the density field (see \eg \cite{Chodorowski2002,Ciecielag2003,Pueblas2009}).

\subsection{Bulk flow, velocity dispersion, and cosmic Mach number}
\label{subsec:bulk_flow}

The bulk flow (BF) is the dipole (second) moment
of the peculiar velocity field, $\vec{v}(\vec{x})$, in a given region of space (volume). 
Non-zero BF reflects a net streaming motion
towards a particular direction in space.
Thus in the continuous limit of the field $\vec{v}$, for
a spherical region with a radius $R$, it will be
\be
\label{eqn:BF-definition-continuous}
\vec{\BF}(R) = {3\over 4\pi R^3}\int_0^R \vec{v}(\vec{x})\textrm{d}^3x\,.
\ee
{Throughout this paper we will interchangeably use BF and $\BF$ to denote the bulk flow amplitude.}

When the velocity field is sampled by a set of $N$ discrete tracers (\eg\ galaxies) then the above integral becomes a finite sum.
{If each individual galaxy is assigned a weight $w_i$, then the 3D bulk flow vector will be}
\be
\label{eqn:BF-def-sum}
\vec{\BF}(R) = {\sum_{i=1}^{N}w_i\vec{v}_i\over \sum_{i=1}^{N}w_i}\,,
\ee
where $\vec{v}_i$ is the peculiar velocity of the $i$-th galaxy. 
The corresponding dispersion (VD) of the peculiar velocities with respect to the averaged bulk flow is
\be
\label{eqn:VD-def-sum}
\vec{\VD}(R)={\sum_{i=1}^N\left[w_i\vec{v}_i-\vec{\BF}(R)\right]^2\over\sum_{i=1}^{N}w_i-1} \,,
\ee
{where the sum of the weights needs to be $\neq1$, so the denominator does not take a zero value.}
If the density fluctuations are a random Gaussian field, then in the linear theory ({\ie{}} on sufficiently large scales)
the corresponding velocity field
will also be a random variable (for each vector component separately) 
with a zero mean and the variance given by the velocity power spectrum
$P_{vv}(k)\equiv\langle v_kv^*_k\rangle$, where $v_k=|\vec{v}_{\vec{k}}|$ and we already assumed global isotropy. 
Thus the predicted {root mean square value of the} bulk flow amplitude
is
\be
\label{eqn:BF_linear_theory}
{\BF}^{2}(R)={1\over 2\pi^2}\int \textrm{d}k k^2P_{vv}(k)|\hat{W}(kR)|^2\,.
\ee
Here $\hat{W}(kR)$ is the Fourier image of the window function. Usually one takes $W$ to be spherical top-hat, which implies
$\hat{W}_{TH}(kR)=3[\sin(kR)-kR\cos(kR)]/{(kR)^3}$, but some authors consider also the so-called all-sky Gaussian selection function
with $\hat{W}_G=\exp(-k^2R^2/2)$.

Now, if there is no velocity bias and the velocity field is {curl-free,} 
then $P_{vv}(k)=k^{-2}H_0^2P_{\theta\theta}(k)$,
and equation \eqref{eqn:BF_linear_theory} becomes
\be
\label{eqn:BF_linear_theory_Ptt}
{\BF}^{2}(R)={H_0^2\over 2\pi^2}\int \textrm{d}k P_{\theta\theta}(k)|\hat{W}(kR)|^2\,.
\ee
In the regime where the velocity vorticity is negligible and the Eqn.\ \eqref{eqn:pk_ptt_equality} holds, one finally obtains
\be
\label{eqn:BF_linear_theory_Pdd}
{\BF}^{2}(R)={H_0^2f^2\over 2\pi^2}\int \textrm{d}k P(k)|\hat{W}(kR)|^2\,.
\ee
The above equation is commonly used as the linear theory prediction for the bulk-flow amplitude in a universe described
by a particular choice of $P(k)$ and $f$. Consequently, the corresponding {variance of the residual velocity field
(after the BF was subtracted) for that case takes the form}
\be
\label{eqn:VD_linear_theory_Pdd}
{\VD}^2(R)={H_0^2f^2\over 2\pi^2}\int \textrm{d}k P(k)(1-|\hat{W}(kR)|^2)\,.
\ee
Now to obtain predictions for the bulk flow amplitude and some significance {intervals,} a model distribution function for peculiar velocities
is needed. This is obtained by noticing that for sufficiently large smoothing {scales,} the distribution for a single velocity
component approaches a Gaussian, thus the distribution for the bulk flow magnitude becomes Maxwellian (see \cite{Bahcall1994,ColesLucchin2002}).
Hence for a velocity {field $\vec{v}(R)$} with rms velocity of $\BF$, this is given by
\be
\label{eqn:BF_distribution}
p(v)\dd v=\sqrt{2\over\pi}\left(3\over\BF^2\right)^{3/2}v^2\exp\left(-{3v^2\over 2\BF^2}\right)\dd v\,.
\ee
Considering $\dd p(v)/\dd v=0$ gives in the limit the most likely value (MLV) $\BF_{MLV}=\sqrt{2/3}\BF$ and the expected value (EV)
$\langle v\rangle=\BF_{EV}=2\BF_{MLV}/\sqrt{\pi}=\sqrt{8/3\pi}\BF$. MLV and EV are widely used as {common linear theory (LT)}
predictions for {the} BF
amplitude, {and} in the reminder of this manuscript we shall adopt the same strategy whenever we will be invoking LT {formulas}. We caution
however, that in this context it is important to bear in mind that such {predictions} only {hold if} the distribution of 
{the} components {of $\vec{v}$}
is Gaussian. The validity of this assumption depends on {scales which} one considers. Although in general 
it was established that for most scales
{dealt with} in modern velocity analysis ({\ie{}} $\simgt 30\hmpc$) 
{this assumption generally holds}
{\cite{Andersen2016}}, results shown by other authors {imply} that caution
should be taken \citep[see also][]{Chodorowski2002,Ciecielag2003}. 

{A separate note  should be made here about the limits of the integrals used to calculate
$\BF(R)$ and $\VD(R)$ from Eqns.~(\ref{eqn:BF_linear_theory})-(\ref{eqn:VD_linear_theory_Pdd}). To obtain predictions for the physical
Universe one should consider the obvious limits from $k_{min}=0$ to $k_{max}=\infty$. However, when we want to compare LT predictions with numerical
simulations that used some finite computation box, we should account for the fact that the modeled velocity field will miss the contribution
from the modes larger than the box length $L$. Also due to discretization of both mass and volume there is some characteristic
minimal scale that is still resolved by the simulation, usually taken to be the force resolution $\varepsilon$. In such a case, the corresponding
integration limits are then ${2\pi\over L}\leq k \leq {2\pi\over\varepsilon}$. Whenever we will be comparing LT predictions with
the simulation results we will employ the above integration limits.}

Some authors \cite{Ostriker1990,Ma2012,Agarwal2013} advocated also another {type of} statistics, 
namely the cosmic Mach number {(CMN{, or $\MN$})}, that we can define now as
\be
\label{eqn:MN-def}
\MN(R)\equiv{\BF(R)\over \VD(R)}\,,
\ee
which in the linear regime should be only a function of the shape of the matter power spectrum (or the effective slope of $\sigma^2(R)$ around $\sim R$)
\cite{Suto1992,Strauss1993,WatkinsFeldman2007}.

\begin{figure}
 \includegraphics[angle=0,width=0.48\textwidth]{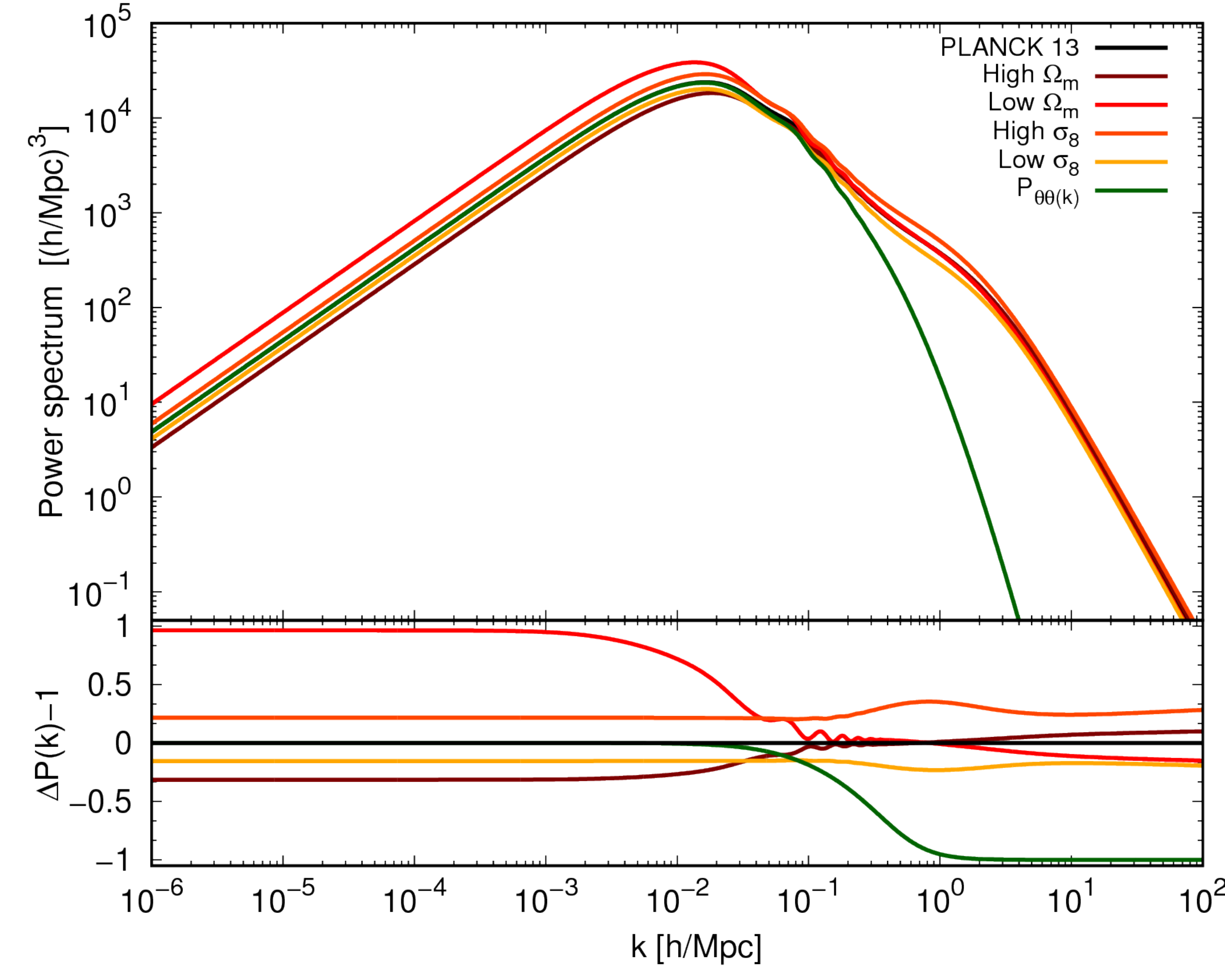}
 \caption{Comparison of the non-linear Planck13 cosmology power spectrum (black solid line) with 
 its variants computed for high and low values of $\Omega_m$ and $\sigma_8$ parameters. In addition,
 the corresponding non-linear velocity $P_{\theta\theta}(k)$ is also plotted with a short-dash-dotted green line. The upper panel
 shows the absolute values, while the bottom panel presents the relative difference with respect to the Planck13 case. 
 }
\label{fig:linear_theory_Pk}
\end{figure}

\begin{figure}
 \includegraphics[angle=0,width=0.48\textwidth]{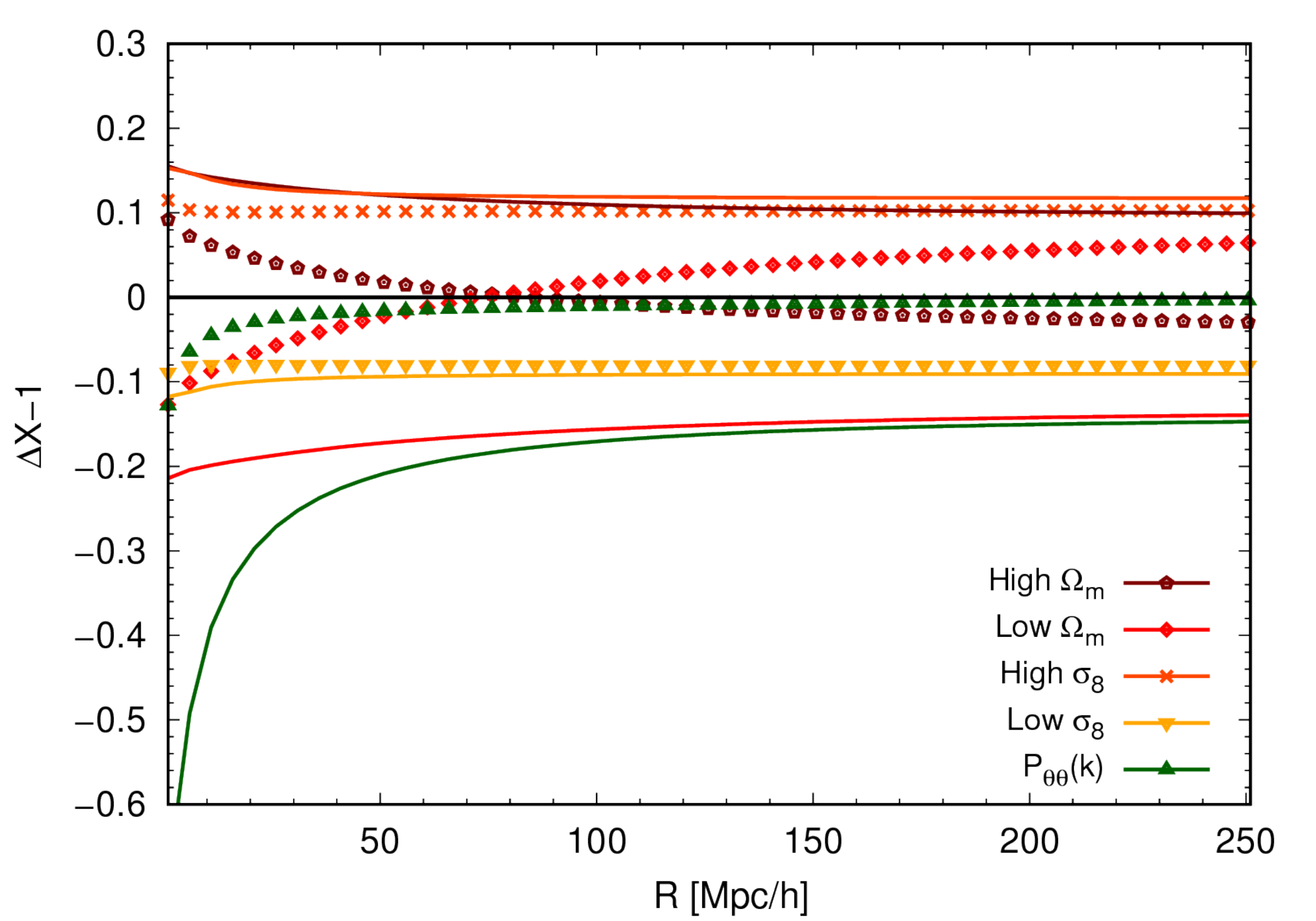}
 \caption{The relative difference of linear theory predictions for the bulk flow magnitude {$\BF(R)$} 
 and the velocity dispersion {$\VD(R)$}
 as predicted by equations (\ref{eqn:BF_linear_theory_Ptt})-(\ref{eqn:VD_linear_theory_Pdd}) taken with respect to 
 the fiducial case where Planck13 non-linear matter power spectrum was used. The symbols mark the results for the bulk flow amplitudes, 
 while the matching color lines are for the corresponding velocity dispersion.}
\label{fig:linear_theory_predictions}
\end{figure}

The above considerations suggest that the linear theory prediction for the bulk flow and associated statistics should strongly depend on
two parameters of the underlying cosmological model, namely the growth rate $f$ and the amplitude of $P(k)$, which can be evaluated by the 
$\sigma_8$ parameter.
These dependencies have motivated many 
authors to advocate the use of the low-order velocity field statistics as
cosmological probes \cite{Strauss1995,Nusser2011b,Branchini2012,Ma2012,Hong2014,Feix2014,Koda2014,Hoffman2015,Scrimgeour2016,Howlett2017}.

To gauge the magnitude of variations and their co-dependence on $f$ and $\sigma_8$ we have considered a number
of power spectra variants. The fiducial case is \textit{i)} the Planck13 cosmology (\cite{Planck13}; the same as used in our simulations) and 
we also examined
four cases: \textit{ii)} high $\Omega_m$ ($\Omega_m=0.35,\Omega_\Lambda=0.65$); \textit{iii)} low $\Omega_m$ ($\Omega_m=0.25,\Omega_\Lambda=0.75$); 
\textit{iv)} high $\sigma_8$ (=0.9) 
and \textit{v)} low $\sigma_8$ (=0.75). Here, for each case \textit{i)--v)} we kept fixed all the remaining 
$\lcdm$ parameters, imposing $\Omega_k=0$ and $\Omega_\mathrm{tot}=1$, and varied only the value of a given matter density or 
power spectrum normalization.
By changing $\Omega_m$ we probe different values of the growth rate (by $\sim10\%$ around the fiducial case) and by 
varying $\sigma_8$ we sample different power spectrum amplitudes. For all the cases we have used the \verb$CAMB$ software package \cite{CAMB} 
to obtain
high-accuracy linear matter power spectra and then applied the \verb$halofit$ model \cite{halofit} to evolve the spectra to the non-linear regime. 
In addition, we also considered one more case, where we used the fully non-linear $P_{\theta\theta}$ estimated
from our $\lcdm$ simulation. The non-linear velocity divergence power spectrum was used only for $k>0.01\mpch$, where it deviates 
by more than $3\%$ from
the non-linear $f^2P(k)$; for smaller $k$ it was substituted by the CAMB-provided $P(k)$, rescaled by $f^2$. We checked the effect of 
the non-linear divergence spectra, since the scales at which the velocity field is curl-free and {the} scales 
at which $\de\ll1$ are not necessary the same \cite{Ciecielag2003,Pueblas2009}.

In Fig.~\ref{fig:linear_theory_Pk} we compare all the examined power spectra with our fiducial Planck13 case \textit{i)}. 
The velocity divergence power spectrum was scaled by the corresponding $f^2$ factor. We can observe that for the cases where $\Omega_m$ is varied,
the corresponding changes in $P(k)$ are limited to large scales, $k\leq 0.1\mpch$. Small deviations seen above $k\simgt 3\mpch$
reflect the different length of the matter-dominated epochs in low and high-$\Omega_m$ universes and so different degree
of non-linearity in the density field. However, this appears at scales too small to be relevant for the large-scale velocity field. 
As expected the high-(low-)$\Omega_m$ case
is characterized by a smaller (larger) amplitude
of the power spectrum than the fiducial case at these scales. For both cases the changes in the large-scale $P(k)$ amplitudes are
quite dramatic. Variations in $\sigma_8$ alone affect the spectrum on all scales, but the {overall} effect is much smaller 
({typically within} $<25\%$).
Here we can also note that the small-scale variance of $P_{\theta\theta}(k)$ is strongly suppressed compared to the matter $P(k)$. 
This is expected, once
one considers that in the non-linear {regime,} while the collapsed objects increase the density field variance, the corresponding 
velocity field around
and inside those objects attains a high-degree of vorticity and dispersion due to shell crossing and virialization
\cite{Pichon1999,Pueblas2009,Libeskind2013,Libeskind2014}.

Figure \ref{fig:linear_theory_predictions} illustrates the changes, imposed due to variations in $P(k)$ shape and amplitude, 
in the corresponding estimated \BF\ (symbols) and \VD\ (lines). The previously seen dramatic differences 
in $P(k)$ amplitudes are translated to rather mild impact on the resulting
linear-theory bulk flows. Here, for the most {cases,} the changes {are within} $\sim10\%$, thus of the same magnitude as our variations {in both}
$f$ and $\sigma_8$. We can also notice the known $\Omega_m-\sigma_8$ degeneracy, where the effect of increasing one parameter can be to a large extent 
compensated by the decrease of the other. The effect of using the non-linear $P_{\theta\theta}(k)$ to predict \BF~ is minimal 
for $R>50\hmpc$. In contrast,
the use of the non-linear velocity divergence power spectrum results in a much more dramatic effects onto the \VD~ estimator.
This suggests that modeling of non-linearities in the density and velocity field is not that important for \BF~\ predictors, 
but might be crucial for the prediction of the expected Mach number. The latter fact was already {emphasized to some extent by \cite{Agarwal2013}},
who noticed that in order to obtain more accurate predictions for the Mach number some non-linear corrections for $\VD$~ have to be
applied. This reflects the fact that the velocity dispersion is intrinsically a local quantity, and non-linear effects such as virialization and
shell-crossing have a significant effect (see \eg \cite{Hahn2013,Hahn2015,Rampf2017}).

\section{The velocity mocks and non-linear observables}
\label{sec:mocks_non_linear}

To move beyond the linear theory we employ the set of $N$-body simulations described in Sec.~\ref{sec:simulations}. 
To study various systematics, non-linear
effects and {biases,} and to get a closer connection with real astronomical {observations,} we construct a set of mock catalogs 
and observables from 
our simulations. As an input for all our analysis we consider halo and subhalo catalogs saved at $z=0$. 

{Generally, when considering various
observational errors and systematics (like survey geometry, selection function, radial distribution, etc.) one can apply their modeling to 
the simulation data and then analyze the mock catalog by computing various statistics from it. We adopt this routine approach by
calculating various data points weights, which characterize different modeled effects in separate mock catalogs.}

We consider the following ``observational effects'' on the data:
\begin{itemize}
 \item {\it observer location} -- all the relevant quantities, such as distances and angles, depend on a specific observer location, 
 whether it would be a random or pre-selected 
 observer; computations are done in the CMB rest frame;
 \item {\it radial selection} -- we model the following radial selections: \textit{1)} full completeness 
 (\ie{} no radial selection nor distance limit); 
\textit{2)} CosmicFlows-3-like \cite{CF-3} selection functions (see below);
 \item {{\it geometry/Zone of Avoidance} -- since} all our catalogs are observer-dependent, it is natural to also include 
 the effect of the so-called
Zone of Avoidance {(ZoA)} caused by obscuration of the far-away objects by the Galactic disk. This is done by removing galaxies from 
the appropriate part of the volume. {See more details below.} 
{In our analysis we do not model the importance of particular structures hidden behind the ZoA, 
such as the Norma Cluster \cite{Norma} or recently discovered Vela Supercluster \cite{Vela}, as this would require detailed 
constrained simulations. We postpone such studies for future work;}

 \item {\it radial velocity error} -- to model {peculiar} velocity errors associated with the uncertainties of galaxy scaling relations
 that are used to infer galaxy velocities from redshifts (see more below). 
\end{itemize}

In our analysis we are concerned with lower-order velocity statistics that are estimated from specific observer-dependent mock catalogs.
Therefore, all our results (unless clearly emphasized otherwise) are computed as ensemble averages over all mock observers
in a given sample. Refs.\cite{LiPan2012} and \cite{Andersen2016} have shown that the distribution of bulk flows amplitudes inferred from 
simulations deviates from a Gaussian. We have checked that this is the case for all our samples, both for the bulk flows as well as for 
the velocity dispersions. For that reason, a simple
averaged mean and associated variance might not be a faithful characterization of the underlying ensemble. Thus we decided to use 
{\it medians} and associated {$16-th$} and { $84-th$ percentiles} to characterize all our results.
 
{\it Observer location}. All the observables we discuss later in the paper were estimated for a fixed given number of observer locations.
By {construction,} all our observers must sit in a DM-halo. We consider two types of observer locations: random and pre-selected. Random
observers are chosen randomly from all halo positions in a given catalog, while the pre-selected are contained in a closed list of
locations predefined by some user provided criteria. In this paper we consider various criteria of a hypothetical 
Local Group (LG)-like observer. See more in Sec.~\ref{sec:LGO}.

{\it Radial selection}. Generally, to obtain the desired radial selection, we would have to select multiple times from an input data
set according to probability that is inversely proportional to the defined shape of the selection function, keeping finally the data
product with mock radial selection that is closest to the imposed one. Such a procedure for large samples as ours is however very unpractical.
We decided to use a simple data weighting scheme instead, where each galaxy is given a weight exactly as defined by the input selection
function. For a large number of galaxies, the results of both procedures give comparable results. Therefore, since we do not
compare our results to any particular galaxy survey, but rather aim at providing general observational data modeling, we are satisfied
with the much faster data weighting method.
{We opt to use weighting scheme that follows the radial selection of the Cosmic-Flows 3 catalog for the sake of simplicity
and generality. CF3 is currently the largest peculiar velocity catalog, 
thus by studying CF3-like radial selection our model will be close to the best-case data scenario.}
In the case \textit{1)} listed above, all the halos have equal unit weights, as in 
Eq. \eqref{eqn:BF-def-sum}.
When modeling the CF3-like radial selection \textit{2)}, we impose \cite{Hellwing2017}:
\be
\label{eqn:data_weighting1}
w_{h}=
\begin{cases}
1, & \text{if}\ r\leq r_w \\
(r/ r_w)^{-m}, & \text{otherwise}\, .
\end{cases}
\ee
Here, $r_w$ is the characteristic radial depth of the catalog (in $\hmpc$). For our CF3-like catalogs we consider $r_w=80\hmpc$ and 
two values for the exponent $m=2,3$;

{\it {Geometry / Zone of Avoidance}.} 
Most extragalactic observations, including those of peculiar velocities, do not have access to low Galactic latitudes due to the obscuration 
by dust, gas, and stars in the Milky Way -- the `Zone of Avoidance'. To model it, we consider a small opening angle $\alpha_{ZoA}=10.5\deg$ 
\cite{ZoA}
chosen with respect to a fixed observer-dependent local $(x,y,z)=(x_{Obs},0,0)$ plane. 
Galaxies falling inside $-\alpha_{ZoA}\leq \alpha \leq \alpha_{ZoA}$ are removed.

{\it Radial velocity error}. Galaxy peculiar velocity surveys rely on redshift-independent {distance-indicators relations (DIs)} to extract 
the cosmological and peculiar components from a galaxy redshift. The most commonly used methods are based on galaxy scaling relations,
such {as Tully-Fisher} \cite{TF} or Fundamental Plane \cite{FP}. Such methods are unavoidably {plagued} with significant
relative errors on estimated velocities stemming from intrinsic scatter in used relations and various systematic 
(usually non-linear) biases. The peculiar velocity errors are a source of a serious worry and their magnitude sets a fundamental 
limit on cosmic velocity data usability. 
A constant relative error in distance determination translates here to a velocity uncertainty that grows linearly with galaxy redshift.
We attempt to model this by a simple relation of the from:
\be
\label{eqn:vel_error_model}
\sigma_v=\Delta_v H_0 D_z\,\,. 
\ee
Here $H_0$ is the Hubble parameter, $D_z$ is the galaxy co-moving distance and $\Delta_v$
models the typical scatter of the logarithmic distance ratio $\eta\equiv\log_{10}(D_z/D_r)$ error. 
The ratio $\eta$ is used to estimate the peculiar velocity.
Here $D_r$ is the co-moving distance to a galaxy inferred via DIs
(see more in \eg{} \cite{Courtois2013,Scrimgeour2016,CF-3}) {and the spectroscopic galaxy redshift $z$}.
We choose $\Delta_v=0.25$, which is
a conservative value when compared with smaller scatter typically found in modern velocity data \cite{CF-3}. 
{We assume that the above velocity error is Gaussian with zero mean and dispersion $\sigma_v$.
In reality such an assumption is often broken for various velocity estimators, but we adopt it
for simplicity, as non-Gaussian contributions to velocity errors depend strongly on particular galaxy catalog 
specifics.}

{Once parameters for mock galaxy catalogs are chosen, we compute the bulk flow and the dispersion
of the residual velocity field by assigning specific halo/galaxy weights and using Eqn.(\ref{eqn:BF-def-sum})-(\ref{eqn:VD-def-sum}).
We sum separately over the three Cartesian velocity vector components in concentric spheres of radius $R$ around a fixed observer location.
This procedure yields us specific weighted bulk flow components, \ie{} $B_x(R)$, $B_y(R)$ and $B_z(R)$. 
The bulk flow amplitude is then
\be
\label{eqn:sum_BF_amplitude}
\tilde{B}(R) = \left(\sum_i^3 B_i(R)^2\right)^{1/2}\,\,.
\ee
Here the sum runs over three Cartesian components of a 3D velocity vector field and the procedure for the residual velocity dispersion is analogous.}

{In reality, the above procedure cannot be applied to real data, since except for a very few cases, 
we do not have full 3D peculiar velocity information. What is directly accessible is only the line-of-sight (l.o.s.) velocity component. 
Thus for observational data one usually adopts an estimator of
the BF that is based on the radial velocity component. For example, in the most popular Maximum Likelihood (ML) method, 
the BF components are obtained via
\be
\label{eqn:sum_BF_ML}
\tilde{B}_i=\sum_n^N w_{i,n}V_n\,\,,
\ee
where $i$ again indicates one of the three Cartesian indexes, $V_n$ is an $n-$th l.o.s. velocity measurement. 
Here, $w_{i,n}$ is an associated
weight of a given velocity measurement, which usually is taken to be
\be
\label{eqn:ML_weights}
w_{i,n}=\sum^3_j A_{ij}^{-1}{\hat{r}_{n,j}\over \sigma^2_n+\sigma^2_*}\,,
\ee
where $\hat{r}_{n,j}$ is a unit vector from the observer to a given galaxy $n$, $\sigma_n$ is the uncertainty of a given
velocity measurement, 
and $\sigma_*$ describes 1D velocity
dispersion due local non-linear virial motions. The matrix $A_{ij}$ describes geometric moments of the whole sample of tracers,
and is given by
\be
\label{eqn:ML_matrix_A}
A_{ij}=\sum_n^N{\hat{r}_{n,i}\hat{r}_{n,j}\over \sigma^2_n+\sigma^2_*}\,.
\ee
The above estimator is based of inverse variance weighting method of Ref.~\cite{Kaiser1988}.} 

{We do not choose to implement the above estimator
for various reasons. First, it is uniquely defined for a given astronomical data set, with its specific radial and geometrical selections and 
errors of velocity estimates. To keep our discussion as general as possible we opt to use a much simpler estimator of Eqn.~\eqref{eqn:sum_BF_amplitude}
instead.
This is justified since all our mock catalogs are isotropic and spatially uniform. For such a case the geometric matrix $A_{ij}$
is uniform and approximates a product of a constant factor and a unit matrix. In addition, since we only use central halos, 
contributions from any non-linear virial motions are strongly suppressed. The last statement does not hold for non-relaxed systems, but 
those constitute a marginal fraction of our $z=0$ halo catalog. Thus we also opt to drop the non-linear velocity dispersion contribution, $\sigma_*$,
from our modeling. }

{Finally, taking into account above considerations and for the sake of simplicity, we choose to use a maximally simplified 
ML estimator, which only 
includes individual velocity errors in the data weights drawn from a Gaussian distribution independently for each velocity component 
according to the prescription of Eqn.~(\ref{eqn:vel_error_model}).} 

\section{Observer-independent systematics}
\label{sec:obs-independent}

Here we will present the results of our analysis of the BF and CMN inferred from mock catalogs where the observer location
was kept random and unspecified, \ie\ it corresponds statistically (after averaging) to a Copernican observer (see more in \cite{Hellwing2017}).
By adopting this approach we will be able to study various systematic effects that are, in principle, independent from the location. 
By doing this we can assess how much the various systematics can affect the measurements in an idealized survey.

\begin{figure}
 \includegraphics[angle=0,width=0.48\textwidth]{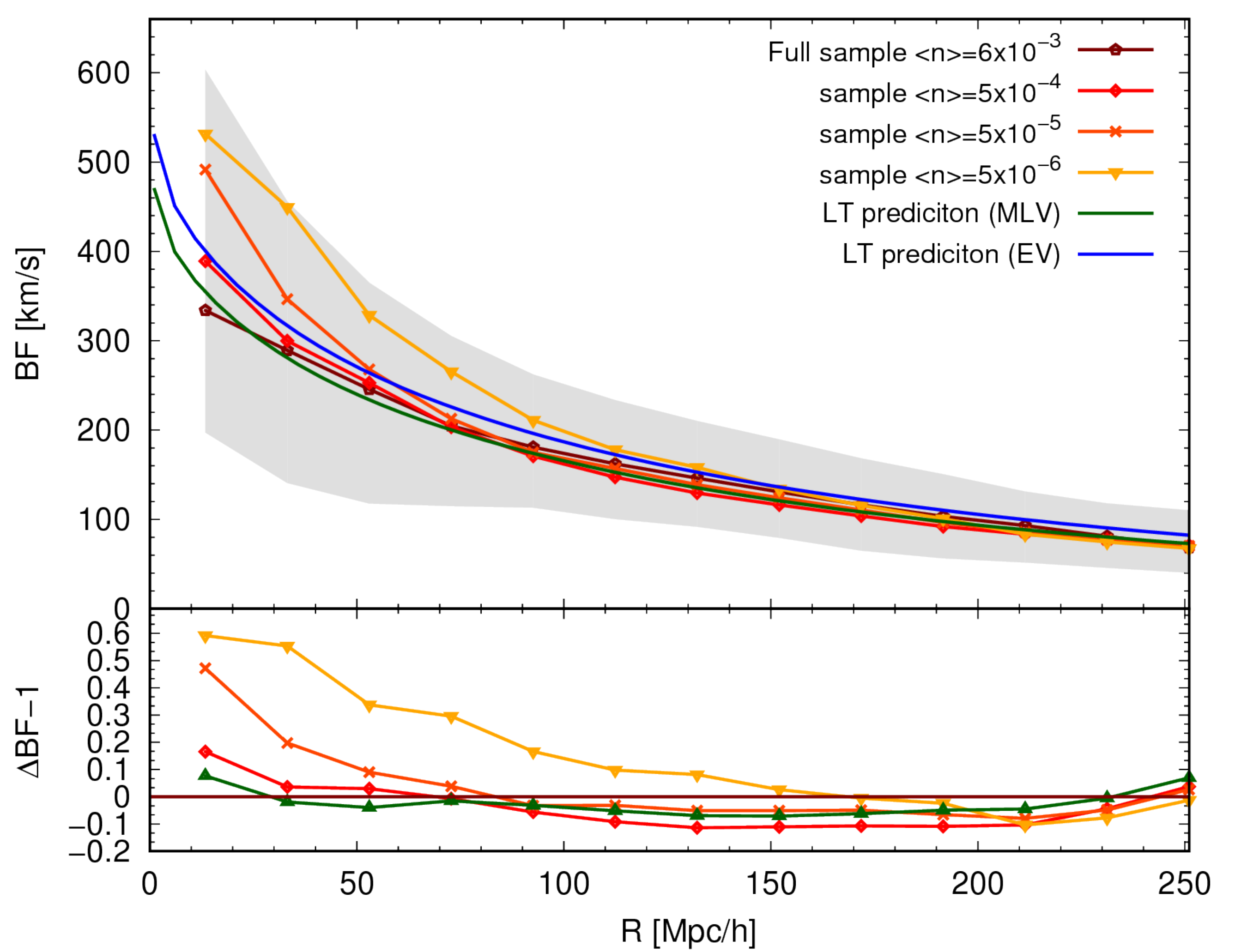}
 \caption{Comparison of bulk flows estimated from {tracer} samples with different number densities.
 {\it Upper panel.} The bulk flow amplitude estimated from simulations (lines with symbols) set together with two linear theory
 predictions: most likely value (green line) and expected value (blue line). The shaded region marks the interval between
 $16$-th and $84$-th percentiles around {the} median value for the full sample (pentagons). Diamonds, crosses and triangles correspond
 to less dense samples of $5\times10^{-4}$, $5\times10^{-5}$ and $5\times10^{-6}$ ($\hmpcc$) respectively.
 {\it Bottom panel}. Relative difference of various tracer samples taken with respect to the full sample result.}
\label{fig:BF_smapling_effects}
\end{figure}

\subsection{Bulk flow}

We begin by investigating how the sampling rate or number density of tracers used for the measurement affects the resulting BF. 
In Fig.~\ref{fig:BF_smapling_effects}
we show the median bulk flow measured for the full sample which is {characterized} by \sam_main{}, and {for} three 
catalogs with lower
number density of tracers, namely $ \langle n\rangle= 5\times10^{-4}$, $5\times10^{-5}$, and $5\times10^{-6}h^3$Mpc$^{-3}$, 
respectively. 
We also plot two {LT} predictions for {the} MLV and EV.
The lower panel of Fig.~\ref{fig:BF_smapling_effects} illustrates 
the relative differences for the various samples, taken always with respect to the fiducial full one, 
which {includes} all the {central} halos. For the scales $\simgt 100\hmpc$ 
all the samples agree with the fiducial one down to $10\%$.
However, at smaller scales we can notice a clear departure of the BF in the lower number density samples from the fiducial case. 
The scale at which such 
deviations start to be noticeable, as well as the magnitude of the effect itself, depend on the number density of objects in the sample. 
The most diluted sample of $\langle n\rangle = 5\times10^{-6}h^3$Mpc$^{-3}$ is at $100\hmpc$ characterized by median BF amplitude 
already higher by $15\%$ than for the full one,
and this grows dramatically
to {$+40\%\sim+60\%$} at $R<50\hmpc$. This discrepancy gets {the} less dramatic the larger number density we consider. For a sample of 
$\langle n\rangle = 5\times10^{-5}h^3$Mpc$^{-3}$, the scale at which the measured BF departs significantly from the fiducial result 
shrinks to $\sim50\hmpc$, but the magnitude still can 
attain quite remarkable $+50\%$ difference at the smallest scales we consider (\ie\ $10\hmpc$).
The sub-sample of one order of magnitude larger 
number density also deviates from the fiducial case, but only at very small scales $\simlt 25\hmpc$, and the relative difference 
reaches $+20\%$ only for the smallest considered radius.

{There is no physical reason for sparser samples to be characterized by larger bulk flow magnitudes. In particular, we expect
that all samples trace the same large-scale regions of a simulated universe. The increase of the amplitude we observe is a purely
statistical effect. Since the BF distribution is not Gaussian, for sparser samples
the shot noise enlarges the width of the BF distribution. This effect combined with overweighted contribution of the outliers 
results in the observed artificial increase of the measured BF amplitude.}
Still, despite the fact that all the differences between the samples are contained within the 16-th and 84-th percentile variation from 
the median of the fiducial one, they are of a systematic nature and if ignored could be a source of a significant BF bias, especially 
at small scales, where in real astronomical surveys the target selection is rather non-uniform. We will discuss 
the implications of these systematic effects 
in the discussion Section \ref{sec:discussion}.

Separately, we note that the LT MLV is a reasonably good prediction for the true BF at nearly all scales probed. 
{This indicates that choosing suitable integral limits for the LT predictors (as discussed earlier) 
allows to properly account for the missing large-scale power.} 
\begin{figure}
 \includegraphics[angle=0,width=0.48\textwidth]{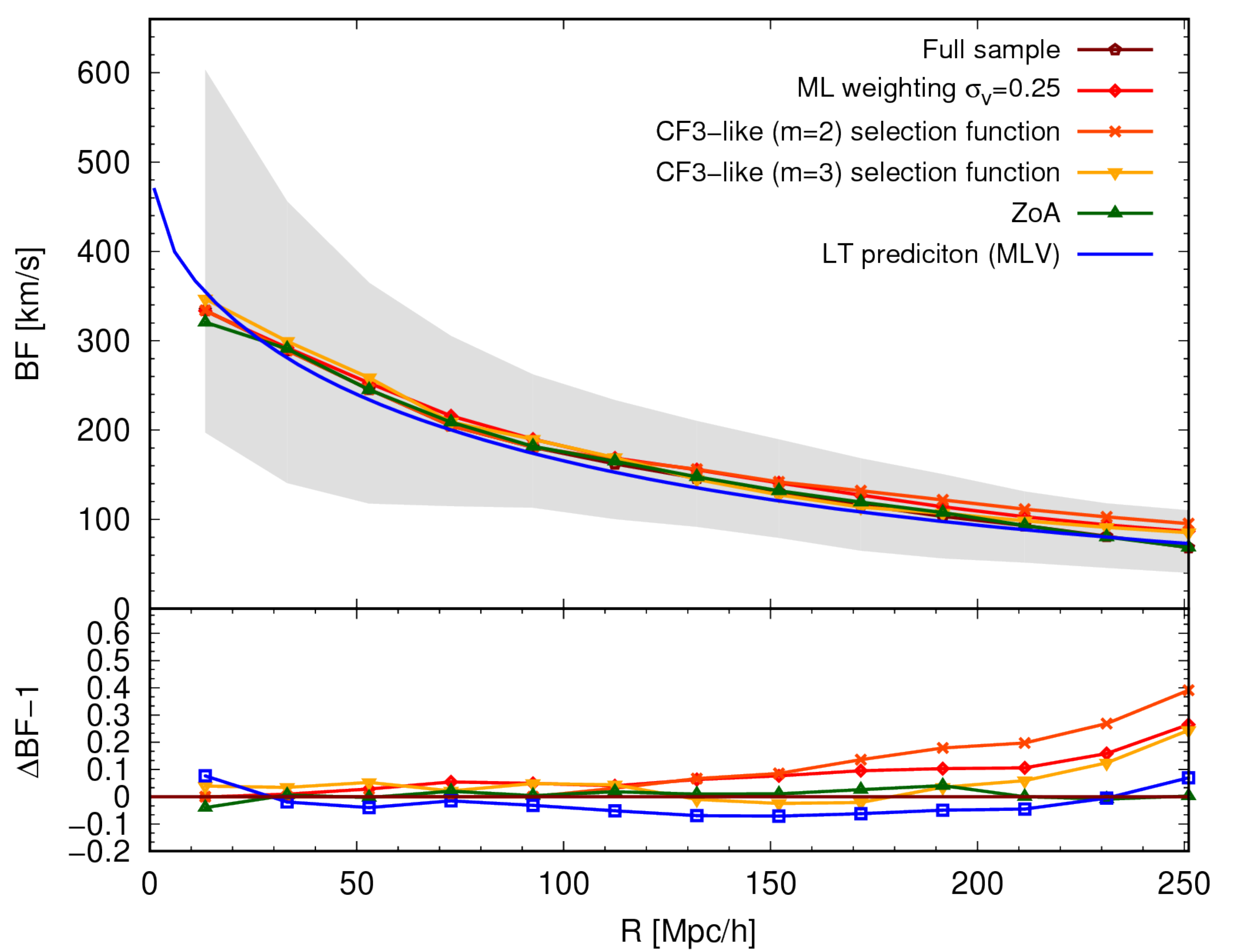}
 \caption{Comparison of bulk flow amplitudes measured from mock catalogs characterized by different observational effects
 considered each separately. The shaded region marks the distance between 16-th and 84-th percentiles around the full sample median (pentagons).
 The comparison is made with velocity error weighting like in the maximum likelihood method (diamonds), CF3-like radial selection with $m=2$
 (crosses) and $m=3$ (down triangles) and Zone of Avoidance geometry (up triangles). For the reference the linear theory prediction
 is also shown (blue continuous line). The bottom panel shows the relative differences taken w.r.t. the full sample values.
 }
\label{fig:BF_weighting_effects}
\end{figure}

We now consider the effects induced on our measured BF by applying various weighting schemes. The galaxy weighting
{prescriptions from Sec.}~\ref{sec:mocks_non_linear} are meant to roughly mimic various systematic effects present in real data. Again, we will
be gauging the measured bulk flow amplitude with respect to our full sample, which constitutes an idealized fiducial case with the best sampling rates 
and no systematics present. For each effect we consider, we apply the specific weighting and data transformation separately
from all the other effects, every time taking the fiducial full sample as a starting point. The situation as presented in our 
Fig.~\ref{fig:BF_weighting_effects} looks quite 
the {opposite to} what was shown in the previous plot \ref{fig:BF_smapling_effects}. 
Here, we observe that the systematic effects (if present) start to matter at large scales
and grow in magnitude with scale. 
The effect related to the observational error modeling as in the ML method is
quite easy to understand, as the error on the velocity grows linearly with scale. This taken together with the Malmquist bias 
\citep{Lynden-Bell1988a,Lynden-Bell1988b}
 {produces} a systematic
overestimation of the measured BF in relation to the full sample \citet{Nusser2014}. The scale dependence of the velocity 
error makes it actually quite easy to model:
for the scales of $R\simlt 120\hmpc$, this weighting overestimates the BF by less than $10\%$. At larger scales $120\simlt R/(\hmpc)\simlt 220$,
it saturates the $10\%$ departure that is rather flat, as no clear scale dependence can be seen. At even larger scales the effect grows up to
$20-25\%$ reaching the maximal expected effect related to the scatter of the intrinsic galaxy relation we use of $\sigma_v=0.25$.

The situation is significantly more complicated for the case of a radial selection function that 
is characteristic of a CosmicFlows-3 like data set.
Here, there is a clear trend that grows systematically with scale, and is related to the effective depth of the sample.
At scales that are above this characteristic depth, $r_w$, which for our case is $80\hmpc$, the BF is grossly overestimated. 
At $R=150\hmpc$ such a radial selection already biases the measurement by $+10\%$ and this quickly grows to values 
of $+40\%$ and larger for $R\simgt 200\sim 250\hmpc$.

When we look at the geometrical selection effect of the ZoA as modeled by us, our results confirm the findings of other authors. 
Namely we find it to have a negligible effect on the measurements, as expected for the case of a symmetric data masking. 
We re-emphasize however that this is valid under the assumption of no significant nearby structures present in the ZoA, 
an effect that we do not investigate in the current paper.

\subsection{Cosmic Mach number}
\begin{figure}
 \includegraphics[angle=0,width=0.48\textwidth]{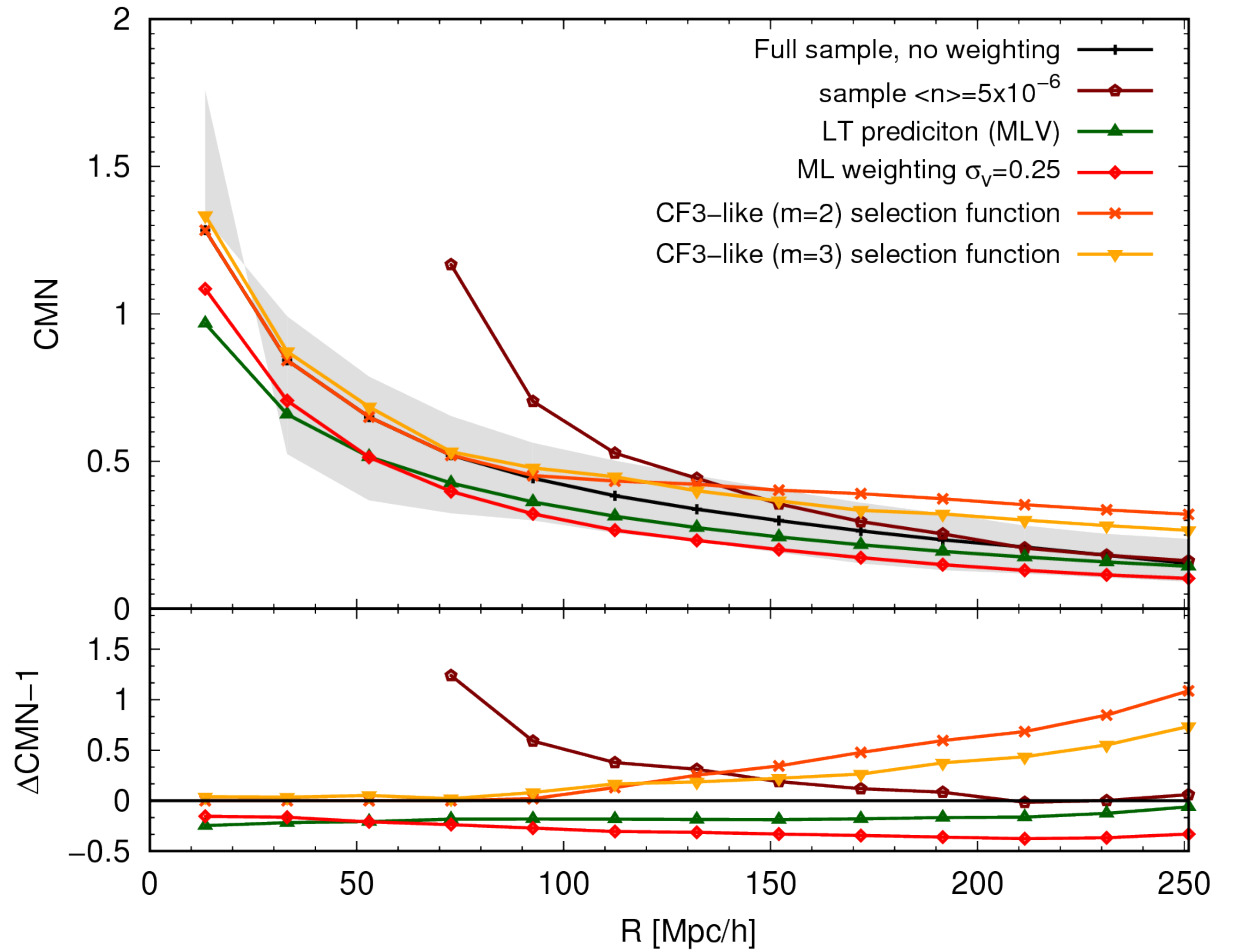}
 \caption{{Analogous} to Fig.~\ref{fig:BF_smapling_effects} and Fig.~\ref{fig:BF_weighting_effects} but for the cosmic Mach number,
 { defined as in Eqn.~(\ref{eqn:MN-def})}.} 
 \label{fig:MN_sampling_anbd_weighting_effects}
\end{figure}

In this paper we analyze also the cosmic Mach number (CMN or $\MN$, interchangeably), 
which, as mentioned earlier, is the ratio of the BF and peculiar velocity dispersion.
The $\VD~$ in a given sphere centered on the observer is not directly observable, however there have been some indirect methods
proposed to measure the CMN \cite{Ostriker1990,Strauss1993,AtrioBarandela2004}.
Thus, we will not present and separately discuss the above mentioned sampling and weighting effects for the VD alone, but rather for
the sake of brevity we show the combined effects on the actual CMN itself. This is presented in Fig.~\ref{fig:MN_sampling_anbd_weighting_effects}.
Again as the reference line we take the fiducial measurement from the full sample. 

The first observation to make is that the magnitude of all visible
systematic effects is significantly larger for the $\MN$~ than it was for the BF. This is not surprising and stems from two facts. First, the VD
is a much more non-linear quantity {than} the BF, as the former strongly depends on short-wavelengths modes; and second, the CMN is a ratio of two
quantities and thus the overall effect of systematic biases and uncertainties is boosted. Moving towards more specific cases, we note that 
{a sparse sample} of $\langle n \rangle = 5\times10^{-6}h^3$Mpc$^{-3}$ leads to a strongly biased $\MN$~ estimate for scales $\simlt 150\hmpc$.
Here, the deviation from
the fiducial case increases with diminishing scale, from $+25\%$ up to more than $+100\%$ bias in a sphere of radius $75\hmpc$. 
We were not able to
probe the CMN for that sample on smaller scales, {since the shot noise from small number counts in such a sparse sample 
dominates there}.
Even for $R<100\hmpc$ we should be careful with interpreting our result, as the mean number of objects in such a volume is 
then $\langle N\rangle<10$.

For the case {of modeled velocity} errors, the estimator clearly provides too low a $\MN$~. We have checked that 
this is a combination of two effects.
Namely, as previously shown, the velocity errors lead to overestimation of the BF, which enters the denominator in the CMN formula. At the same time
the velocity errors naturally lead to underestimation of the local $\VD$. These two combined effects 
make such a CMN estimator, mimicking real data properties,
significantly biased at all probed scales.

The situation {becomes} even more severe for the case of CF3-like radial selection functions. Here,
both examined selections offer highly biased $\MN$~ estimators for all scales larger than the characteristic survey depth $r_w$, and the systematic
effects quickly become catastrophically large. At $R\simlt r_w$ the estimated CMN is very close to the fiducial case; this is not a surprise,
as here the radial selection is still complete (\ie{} is equal to unity). 
Finally, it is also important to note that for the case of the $\MN$~, the LT predictor does not offer
a reliable estimator. This is clearly shown in Fig.~\ref{fig:MN_sampling_anbd_weighting_effects}: LT significantly underestimates the 
CMN for all the considered scales. As we have already
assessed that the LT offers a reasonably good prediction for the BF, we then conclude that it must be the VD which is {underestimated.}
Indeed, this is clearly the case, as was already hinted at by the results shown in Fig.~\ref{fig:linear_theory_predictions}. 
 
\section{Biases for Local Group observers}
\label{sec:LGO}

In this section we will test and quantify {potential biases} that arise in the measurements of the lowest moments of the peculiar 
velocity field if one neglects the fact that the related observations available to
us come from a specific location in the Universe. In other words, we will compare ensemble medians of the low-order moments of 
the galaxy velocity field measured by unspecified observers, whom we will call \textit{random observers} {(\RNDO)}, and different observers placed
at specific locations which fulfill various criteria we consider to be related to {the position} of a Local Group (LG) analog observer. 
Work of Ref.~\cite{Hellwing2017} have shown that such \textit{LG observers} (\LGO)
can exhibit highly biased local velocity correlation measurements.

To stay consistent with this previous study we will consider exactly the same selection criteria used to define a set of LG-analog observers.
For clarity we give here all the essential information, referring the reader looking for more specific details or discussion to the original
work.
The LG is a gravitationally bound system of a dozen major galaxies with the Milky Way (MW) 
and its neighboring M31 as the major gravitational players.
The region of 5 Mpc distance from the LG {barycenter} is characterized by moderate density 
\citep[see \eg][]{Tully1987,Tully1988,Hudson1993,Tully2008,Courtois2013}
and a quiet flow \citep{Sandage1972,Schlegel1994,Karachentsev2002,Karachentsev2003}. 
Located at a distance of {$\sim 12\hmpc$} is the Virgo cluster,
whose gravitational effects extend to tens of Mpcs around us, as evident from the corresponding 
infall flow pattern of galaxies \citep{TullyShaya1984,Tammann1985,Lu1994,Gudehus1995,Karachentsev2014,Libeskind2015}.
The presence of such a large non-linear mass aggregation
can {and does have substantial} impact on peculiar velocity
field of the local galaxies.

To find locations of prospective LG-like observers we use the following criteria:
\begin{enumerate}
\item \label{item:MW} the observer is located in a MW-like host halo of mass $7\times10^{11}<M_{200}/(\Msun)<2\times10^{12}$ 
\citep{Busha2011,Phelps2013,Cautun2014,GuoCooper2015},
 \item \label{item:v} the bulk velocity {(of smoothed DM velocity field)} 
 within a sphere of $R=3.125\hmpc$ centered on the observer is $V= 622 \pm 150\kms$ 
 \citep{Kogut1993},
 \item \label{item:d} the mean density contrast within the same sphere is in the range of $-0.2\leq\delta\leq 3$ 
 \citep{Karachentsev2012,Elyiv2013,Laniakea},
 \item \label{item:virg} a Virgo-like cluster of mass $M=(1.2\pm 0.6)\times10^{15}\Msun$ is present 
 at a distance $D=12\pm4\hmpc$ from the observer \citep{Tammann1985,Mei2007}.
\end{enumerate}
To examine the role of individual criteria we also study results for sets of observers selected 
 without imposing all {the above} constraints. The sets of observers we consider are: 
\begin{description}
\item[\LGO0] a set of the most constrained 2294 observers, each satisfying all the selection 
criteria \ref{item:MW} through \ref{item:virg};
\item[\LGO1] consists of 5051 candidate observers without the velocity constraint \ref{item:v}, 
but satisfying the remaining criteria \ref{item:MW}, \ref{item:d} \& \ref{item:virg}; 
 \item[\LGO2] {includes} 4978 candidates without the density contrast condition 
 \ref{item:d}, but with \ref{item:MW}, \ref{item:v} \& \ref{item:virg};
\item[\LGO3] of 4840 candidates with the conditions \ref{item:v} \& \ref{item:d}
relaxed simultaneously, \ie\ meeting \ref{item:MW} \& \ref{item:virg};
\item[\LGO4] {a set of} 6245 observers without imposing the constraint on the host halo 
mass \ref{item:MW}, but with all the other criteria \ref{item:v}--\ref{item:virg} fulfilled;
\item[\LGO5] contains 288424 candidate observers satisfying the conditions \ref{item:MW}--\ref{item:d} but not the 
proximity to a Virgo-like cluster \ref{item:virg}; 
\item[\RNDO] is a list of observers with randomly selected 
positions in the simulation box. This set is used as a benchmark for comparison. 
\end{description} 
Since the number of prospective candidates in each set is large, to keep the sampling noise at the same level and also to
speed up the calculations we will only consider 125 observers from each set. Since positions of observers are not independent {of each other},
we subsample the candidates by placing a $5\times5\times5$ grid in the simulation box and {keeping only} 
one unique observer location
within each grid cell. All the results shown in this section were obtained by taking {the} median of 
the distribution for all {the} 125 observers in each set. 

\subsection{Bulk flow}
\label{subsec:bf}

\begin{figure}
 \includegraphics[angle=0,width=0.48\textwidth]{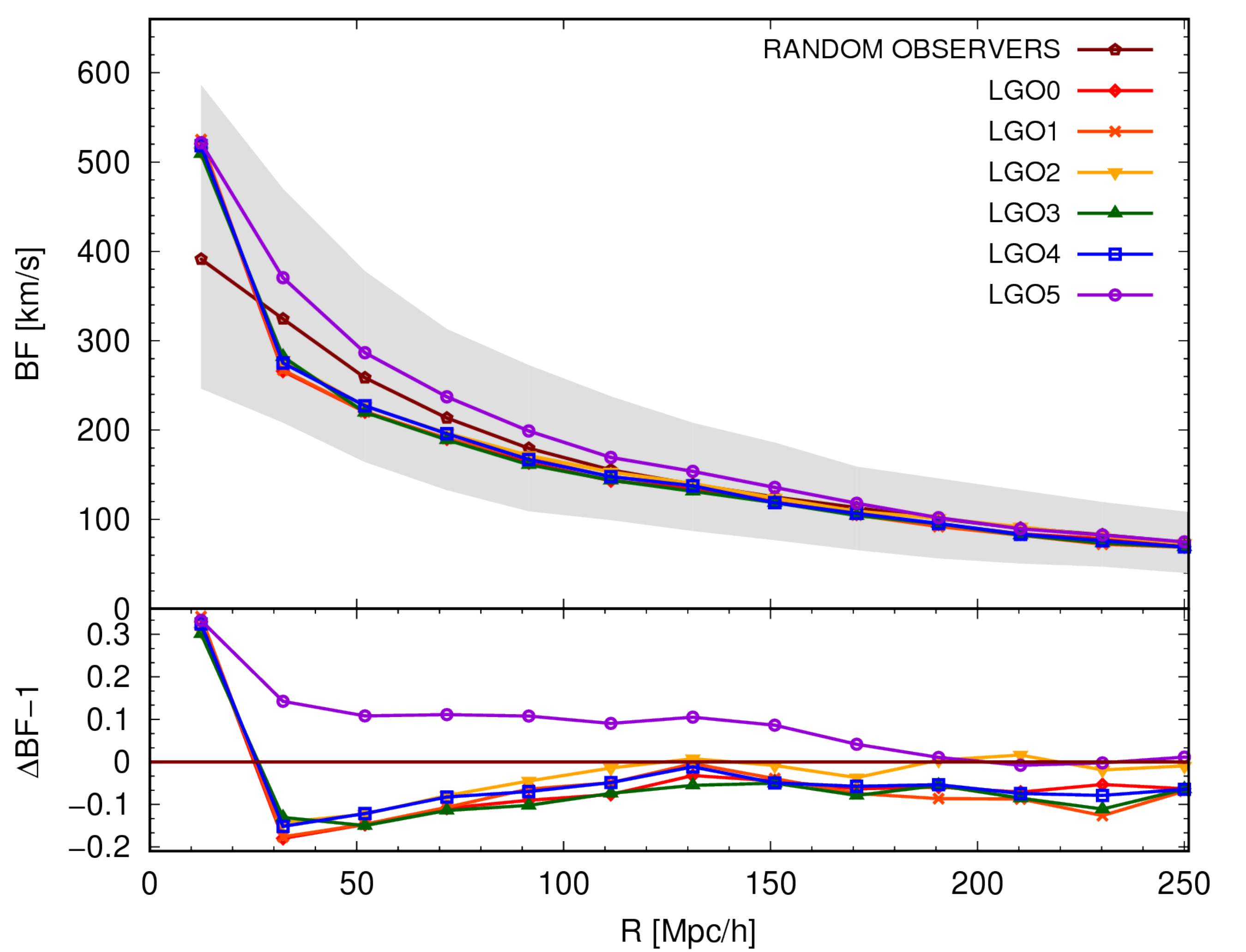}
 \caption{Comparison of the median bulk flow measured by a random observer against those inferred for various Local-Group-like observers 
 (see text for details). {\it Upper panel.} The amplitude of the median bulk flow measured for 
 an ensemble of observers of a given class. {\it Bottom panel.} Relative differences taken with respect to the fiducial random observer case.}
\label{fig:BF_LGO_effects}
\end{figure}

Figure~\ref{fig:BF_LGO_effects} illustrates the systematic effects on the median BF as measured by various observers. As {the} reference
we take the Copernican observer of an unspecified location. 
In other words, we expect that the \RNDO{}
observers measure the expected cosmic mean values. Indeed, the results shown in the previous section agree with this assumption, as the BF measured
for the random observers agrees well with the LT prediction (Fig.~\ref{fig:BF_smapling_effects}). The shaded region in 
Fig.~\ref{fig:BF_LGO_effects} again illustrates the width of the distribution of measured bulk flows
between the 16-th and the 84-th percentiles.

A quick look at the results for different non-random observers already allows us to find a striking feature: there is only one criterion 
really discriminatory for the results. Namely, what matters here is the proximity of a Virgo-like cluster
to the observer. All \LGO{} analogues who fulfill the latter requirement
measure a BF that is systematically smaller than the cosmic mean for $R\simlt 125\hmpc$. 
This effect is around $\sim10\%$ at $\sim100\hmpc$ and grows
to even $20\%$ for scales smaller than $50\hmpc$. Additionally, we see that the LG position requirements considered without the proximity of
a Virgo-like analogue also have an effect on the measured BF. Interestingly, this seems to work in the opposite direction than 
the other joint criteria, and an LG-analogue but no-Virgo
observer would measure actually a systematically larger BF than a random one. This means that the effect of 
the Virgo-like object proximity is actually stronger than shown by our LG-analogues. We have used a small set of 
observers with just the Virgo-criterion to check that this is indeed the case. 

{We propose the following interpretation of these findings.} {The criterion that an observer should be located nearby 
a massive structure of a Virgo-like mass induces a constraint on
the local density (hence also velocity) field when compared to a fully random observer. Such a constraint naturally
lowers the scatter among observers \citep[][]{Hoffman1992,Weygaert1996}, thus also the BF magnitude.}

\subsection{Velocity Dispersion and Cosmic Mach number}
\label{subsec:CMN}

\begin{figure}
 \includegraphics[angle=0,width=0.48\textwidth]{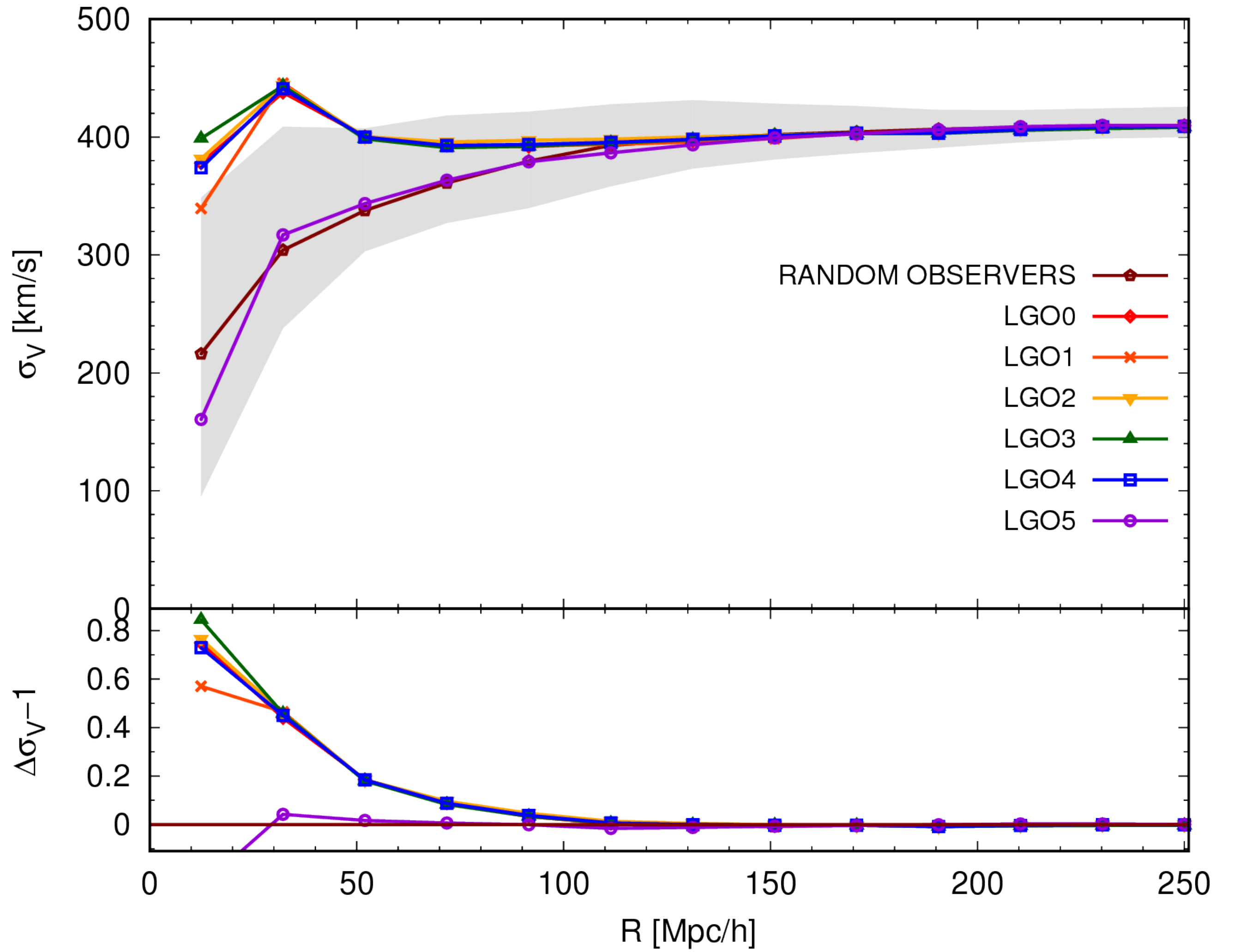}
 \caption{Comparison between the median velocity dispersion measured by a random observer against those for specific Local Group Observers. 
 Analogous to Fig.~\ref{fig:BF_LGO_effects}.}
\label{fig:VD_LGO_effects}
\end{figure}

\begin{figure}
 \includegraphics[angle=0,width=0.48\textwidth]{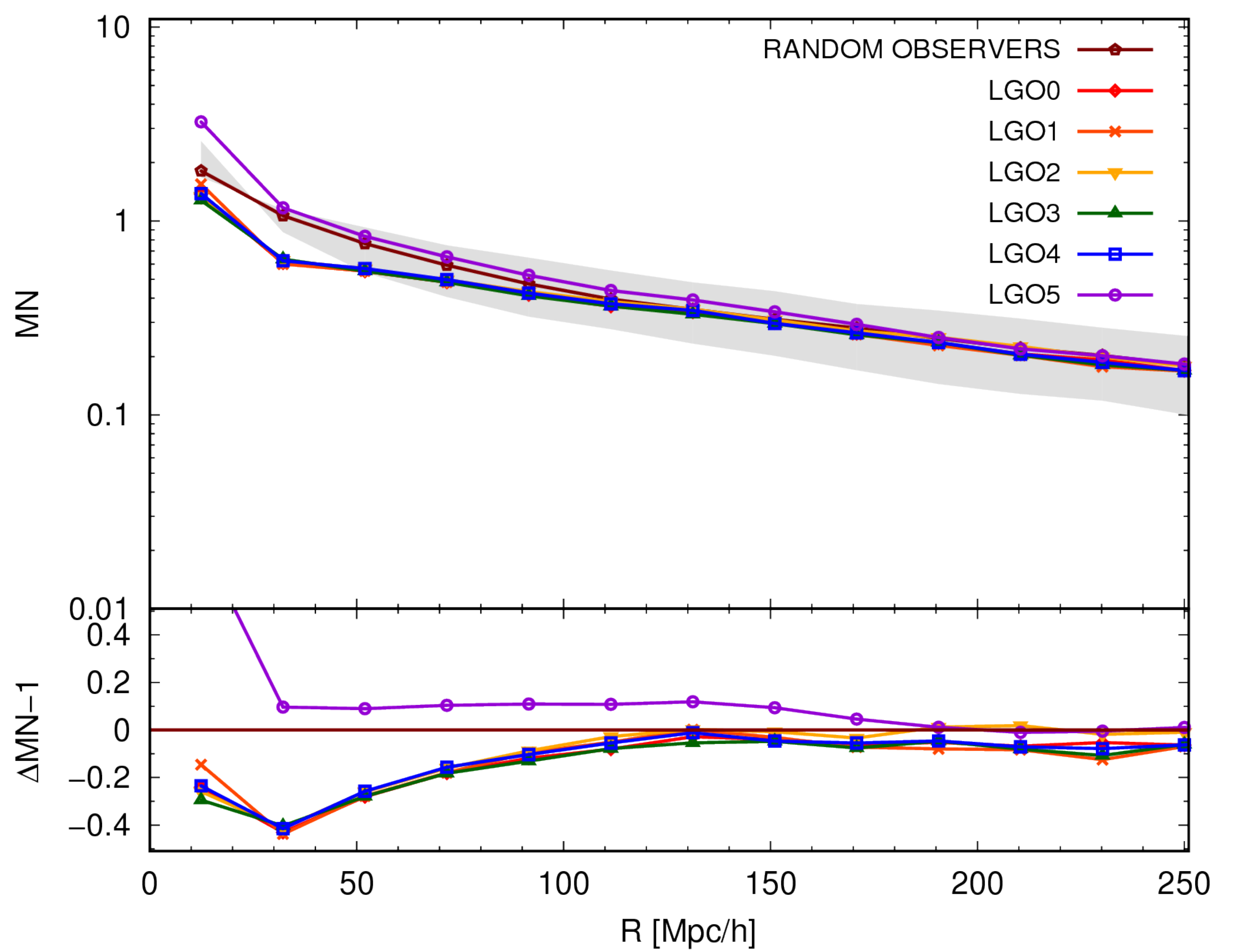}
 \caption{Comparison between the median cosmic Mach number measured by a random observer against those for specific Local Group Observers. 
 Analogous to Fig.~\ref{fig:BF_LGO_effects}.}
\label{fig:MN_LGO_effects}
\end{figure}

We now turn to the importance of observer location for the VD and $\MN$~ statistics. 
In Fig.~\ref{fig:VD_LGO_effects} we plot the comparison of median velocity dispersions obtained for 
the different observers we consider. Here, we notice that the effects imposed by a Virgo-like proximity are contained to somewhat smaller 
($\simlt90\hmpc)$ scales {than} in the BF case. All our LG-analogues with a nearby cluster measure much higher $\VD~$ (up to $50\%$) at small scales.
This clearly indicates that {the} effect is purely driven by the presence of a massive non-linear structure of the cluster. 
Interestingly however, all the measurements converge to the random value at $R\sim110\hmpc$.

The effects of the observer location for the CMN statistics are illustrated in Fig.~\ref{fig:MN_LGO_effects}.
Not surprisingly, it is clear {that the} overall \LGO{} effect is driven mostly by the presence or absence of a nearby Virgo-analogue cluster.
This amounts to \LGO{} $\MN$~ bias of the order of $\sim40\%$ at $R\simlt50\hmpc$, which reduces to $\sim10\%$ at $100\hmpc$. 
Thus, in the case of CMN one is concerned with an even stronger observer bias than in the BF case. This should be remembered and 
accounted for before any cosmological analysis of this statistics is performed.

\section{Gravity and growth rate}
\label{sec:gravity}
\begin{figure}
 \includegraphics[angle=0,width=0.48\textwidth]{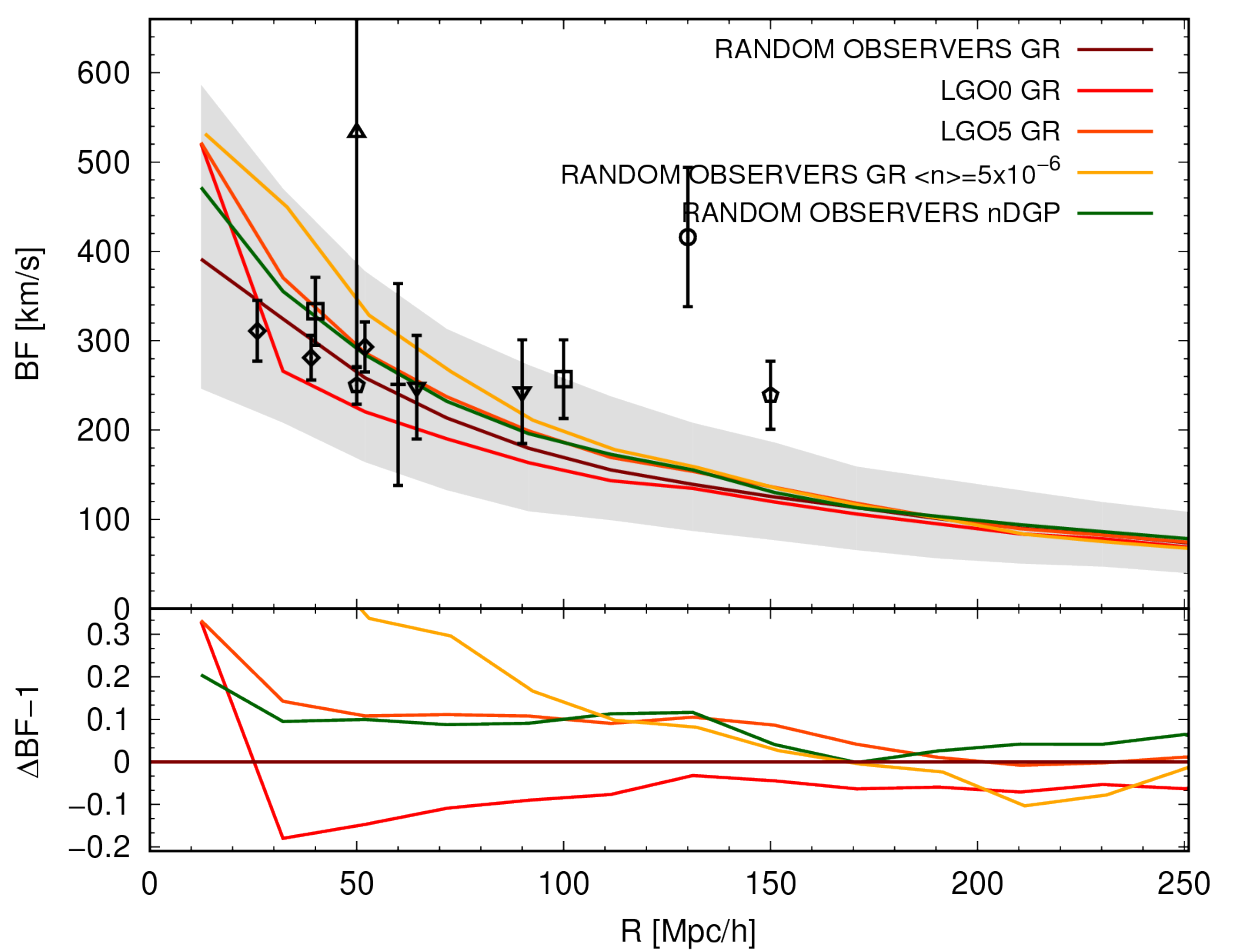}
 \caption{The median bulk flow as measured by random observers in GR and nDGP gravity model compared with GR-observer measurements affected 
 by various systematics. The data points illustrate some recent $\BF(R)${} measurements from the literature:
 dash - Branchini {\it et al.}{}\citep{Branchini2012}, pentagrams - Hoffman {\it et al.}{}\citep{Hoffman2015}, 
 diamonds - Hong {\it et al.}{}\citep{Hong2014},
 down triangles - Scrimgeour {\it et al.}{}\citep{Scrimgeour2016}, up triangle - Lavaux {\it et al.}{}\citep{Lavaux2013}, 
 squares - Nusser\&Davis\citep{Nusser2011a}
 and circle - Feldman {\it et al.}{}\citep{Feldman2010}.}
\label{fig:BF_MG_effects}
\end{figure}

{In the previous sections we have observed that various systematics can significantly change the bulk flow amplitude
on wide range of scales. This fact has fundamental consequences for all applications that hope to use the BF
and related statistics such as $\MN$~ to search for non-GR signature. As an illustration, here we will compare a velocity signal 
from a modified gravity model
against the various observational effects present in the GR case.}

For our {\it guinea pig} {MG} model we choose the normal branch of Dvali-Gabadadze-Poratti (henceforth nDGP) model 
\cite{SahniShatanov2003,LueStarkman2004,Schmidt2009b},
which implements the non-linear fifth-force screening (the {Vainshtein mechanism) \cite{Vainshtein}}
and can be characterized at large-scales by a nearly constant (\ie{} scale-independent) enhancement of the growth rate of structures 
(see also \cite{Barreira2013}).
Specifically we choose to take the value of the so-called crossing-over scale to be $r_cH_0=1$. This value 
represents the scale at which gravity becomes {5-dimensional in this model}. The smaller this scale, the stronger deviations from GR-dynamics (due to
the fifth-force) can be expected. Our choice of $r_c$ gives moderate modifications to GR that are characterized by a linear 
growth rate
(the logarithmic derivative of linear density growing mode) {$f_{nDGP}\approx 1.15f_{GR}$} \cite{Koyama2006,Li2013,Bose2017}.
{Except for the modified dynamics induced by the scalar field present in the nDGP model, our MG simulation} shares exactly 
the same set-up and parameters
as the fiducial GR-case. For the sake of speeding-up the numerical computations we have employed the Truncated DGP method
described in details in \cite{truncatedDGP}. The speed-up is obtained at the expense of the resolution of the scalar-field spatial fluctuations,
solving of which was truncated beyond the 4-th mesh refinement level. This sets the resolution of the scalar force at $\sim60\hkpc$,
which is still considerably smaller {than} the smallest halos we consider. As we use the same initial conditions {for both GR and nDGP,} 
the large-scale cosmic variance effects
should be of the same magnitude in both {runs (see} also \cite{Hellwing2017b}), and the observed discrepancies should reflect 
the differences in the underlying gravitational dynamics.

Figure \ref{fig:BF_MG_effects} compares the BF measured by two Copernican observers, one in GR and one in the nDGP model (marked as MG)
{versus} the amplitudes expected in {the} GR case {with} different {systematic} effects. For the sake of brevity, we choose to compare with only 
the strongest systematics elucidated in the previous Section. In particular, we show the \LGO0{} and \LGO5{} signals, as well as 
\RNDO{} observers with sparse sampling of $\langle n\rangle=5\times 10^{-6}$h$^3$Mpc$^{-3}$. {For $R\simlt200\hmpc$,}
the MG bulk flow is enhanced {by $\sim10\%$}, as one can expect from {the} LT prediction of Eqn.~\eqref{eqn:BF_linear_theory_Pdd}.
This potentially observable effect can be easily obscured by various {systematics that} have larger magnitudes on the same scales. 
Specifically, we see that realistic modeling of the Local Group analogue observers, which includes the effects of the Virgo cluster proximity, 
gives opposite sign to the MG enhancement. Thus, in the worst case scenario, we could have a conspiracy, where a BF signal for a \LGO0{} 
observer in an MG universe would look like a BF expected for a \RNDO{} observer in the GR universe. On the other hand, the signal expected for 
a \LGO{} observer modeled without a Virgo-like cluster presence can mimic {the} scale-dependence and amplitude of a \RNDO{} MG signal.
For a very sparse sample, these two observations would be dwarfed by a systematic effect that on small scales ($R\simlt 100\hmpc$)
can be by a factor of a few times larger than what we can expect for a reasonably mild MG model enhancement.

{The main merit of our work here is to systematically study potential biases of low-order velocity measurements,
but it is illustrative to compare the scales and amplitudes of the effects we report with some $\BF(R)${} measurements 
reported in the literature. We have selected arbitrarily seven such measurements and marked them in Fig.~\ref{fig:BF_MG_effects}.
We show results from Branchini {\it et al.}{}\citep{Branchini2012}, Hoffman {\it et al.}{}\citep{Hoffman2015}, Hong {\it et al.}{}\citep{Hong2014},
Scrimgeour {\it et al.}{}\citep{Scrimgeour2016}, Lavaux {\it et al.}{}\citep{Lavaux2013}, Nusser\&Davis\citep{Nusser2011a} 
and Feldman {\it et al.}{}\citep{Feldman2010}. The methods and datasets used in these references vary significantly, so this collection
is a fair representation of approaches and data used currently in peculiar velocity studies. Except for Refs.~\citep{Hoffman2015}
and \citep{Feldman2010}, all the results are consistent within $16$-th and $84-$th percentiles with median \BF{} of random
and \LGO{} observers and even with the MG model. If we took the size of the error bars reported by those authors
at their face-values, some of our results (such as the MG model) would be marginally inconsistent with that data.
However, we clearly see that the variance added by the systematic effects will boost the reported error bars significantly.}

The results shown here
can have potentially profound repercussions, as it would seem that the lower-order velocity statistics are plagued 
by potentially overwhelming systematic effects that can completely obscure even relatively strong ($\sim10\%$) deviations 
from the GR case. We shall discuss the implication of these findings in the next section.

\section{Summary}
\label{sec:sumarry}

In this paper our main aim was to methodically check various possible systematic effects that could affect the measured values
of the bulk flow, {peculiar} velocity dispersion, and cosmic Mach number. {Peculiar velocities of galaxies} strongly
depend on the underlying cosmic parameters,
such as the logarithmic growth rate ($f$) and the non-relativistic matter energy density ($\Omega_m$).
Velocity data are prone to large uncertainties stemming from the intrinsic scatter of various empirical galaxy scaling relations 
used to measure redshift-independent distances, and consequently, infer peculiar velocities. The latter are additionally 
affected by such issues as non-Gaussian errors and non-linear ({\it i.e.} virial) contributions. 
{Many methods have been proposed and implemented to deal with these issues.}
However, once the velocity data had been corrected for intrinsic errors and Malmquist biases, it was commonly assumed that 
the relevant statistics could be directly related to
the underlying cosmology using theoretical modeling (such as linear perturbation theory). This assumed advantage was
one of the main arguments for using the galaxy velocities as alternative cosmological probes. In our analysis we have revisited this assumption,
and our results indicate that there are many systematic effects that {need} to be accurately modeled and accounted for, 
in order to infer cosmological parameters from low-order velocity statistics in an unbiased manner.

{Below we summarize and comment on all our important findings and their implications.}

\begin{itemize}
 \item {\it {Perturbation} theory estimators:}
 \begin{enumerate}
 \item The results encapsulated in {Fig.~\ref{fig:linear_theory_predictions} show} that strong {modulations} of the {density} power spectrum
 amplitude {lead} to only mild variations in BF and VD. More precisely, we have found that the changes in both statistics are roughly proportional
 to changes {in} $f$ and $\sigma_8$.
 \item Using the non-linear velocity divergence power spectrum instead of {the} non-linear density $P(k)$ has a strong effect on {the} predicted $\VD$,
 and therefore
 also on the {$\MN$~}. This is because the magnitudes of the effects induced in the non-linear regime are opposite in those two spectra. 
 Namely, at small
 scales the non-linear $P_{\theta\theta}(k)$ takes smaller values {than} the linear theory prediction, while the non-linear $P(k)$ has actually
 {a} boosted amplitude w.r.t.\ the linear theory. At small scales in the non-linear regime the motions of galaxies {lose} the character
 of a potential flow. This {reflects significant} growth of vorticity and velocity dispersion due to shell-crossing 
 {\citep{Pichon1999,Pueblas2009,Libeskind2014,Cusin2017}}.
 For that reason it is important to take into account and model properly these non-linear effects, in order to get 
 a more realistic perturbation theory
 prediction for the {CMN}.
 
 \end{enumerate}
 \item {\it Sampling rate effects:}
 \begin{enumerate} 
 \item At small scales ({\it i.e.} $\simlt 50\hmpc$) we found a significant effect on BF magnitude from under-sampling. 
 For increasingly diluted samples the inferred BF is biased towards higher values when compared to our fiducial full sample, 
 and the effect is the larger, the smaller sample density. {This is a statistical effect induced by the non-Gaussian distribution
 of BF magnitudes and increased shot-noise contribution due to sparse sampling.}
 For the sparsest sample with $\langle n\rangle \geq 10^{-6}h^3$Mpc$^{-3}$, the bias could attain $50\%$ or more.
 This indicates that one should {carefully account for this effect} when low-order velocity statistics
 {are} measured {from very} sparse samples like supernovae 
 \cite{Weyant2011,Turnbull2012,Feindt2013,Huterer2015}. 
 \item The down-sampling of the data we have performed is essentially equivalent to weighting the data by the local value of the density
 correlation function,
 since galaxies are biased tracers and form preferentially  in higher-density regions. Such regions 
 are characterized by higher values of local bulk flows.
 \item Our results show that the MLV estimator from linear perturbation theory is {in good} agreement with the $N$-body data for 
 a random-location BF observer.
 \end{enumerate}
 \item {\it Selection effects:} 
 \begin{enumerate}
 \item Weighting halo velocities according to velocity errors (like in the ML method) leads to overestimation of the BF.
 This effect grows {with distance} from the observer, but it saturates to a maximum value that is close to the considered typical scatter
 of the {distance indicator error} $\sigma_v$. 
 \item For a {radial selection} function of a CF3-like survey form, it is evident that the limited depth of {the} catalog is 
 reflected in the measured BF.
 The BF is {formally} an integral over a sphere, but {for a realistic survey of discrete galaxies, it becomes a sum over 
 concentric spherical
 shells of growing radius. The radial selection} function is effectively down-weighting the outer shells, and thus the BF value 
 {derived} from inferior spheres
 is spuriously {propagated to larger scales}. This is a strong effect, which indicates {that BF} measurements {from} 
 scales comparable or larger {than the characteristic depth of a given catalog} should be {interpreted with care.} 
 \item A symmetric sky-map angular incompleteness {with} an opening angle of $\sim10\deg$ {-- such as the Zone of Avoidance --} 
 has a negligible effect on {the} inferred BF.
 \item All systematic effects related to galaxy selection are much more pronounced in the CMN {statistic. 
 In particular, both sparse sampling and} velocity errors induce a significant $\MN$~ bias for scales $\simlt 150\hmpc$.
 \item A radial selection function of the CF3-like form induces a catastrophically large CMN over-prediction for scales larger
 than {the} given survey characteristic depth.
 \end{enumerate}
 \item {\it Observer location effects:}
 \begin{enumerate}
 \item For all our low-order statistics the most {important} \LGO{}-analogue criterion is the proximity of a Virgo-like cluster.
 \item Local Group-analogue observers measure systematically smaller BF amplitudes {than} the cosmic mean (\ie{} {a} random observer)
 on scales up to $R\simeq 125\hmpc$. This systematic {attains $\sim10\%$ at $\sim10\hmpc$ and grows to} $\sim20\%$
 for $R\simlt50\hmpc$.
 \item {A} no-Virgo observer (\ie{} \LGO5) at the same time exhibits an opposite bias, inferring a BF that is larger by $\approx10\%$ on 
 similar scales {than $\BF$ measured by the \RNDO{} observers.}
 \item For the VD the LG-observer bias is contained to somewhat smaller scales $R\simlt90\hmpc$, but its magnitude reaches a quite
 dramatic value of up to $50\%$. This effect is purely driven by {the} proximity of a Virgo-analogue, as {a} no-Virgo observer
 {measures VD values compatible} with $\RNDO$ observers. {This indicates} that an infall region around {a} massive cluster is significantly heating
 up the local velocity field.
 \item The effects observed for BF and VD combine {into a} biased $\MN$~ for \LGO{} {observers, which} manifests itself as $\sim40\%$ 
 under-evaluation at $<50\hmpc$, and still takes $\sim10\%$ too small a value at $R\sim100\hmpc$.
 \end{enumerate}
 \item {\it Growth rate and gravity}
 \begin{enumerate}
 \item The effect of an increased growth rate $f$ {observed in a representative} MG model {is} degenerate with the specific bias 
 induced by a \LGO{} observer{, and the Virgo-like cluster proximity in particular.}
 \item {Similarly,} a non-Virgo {\LGO{}-like} BF signal appears very similar to {the} MG signal for {a} \RNDO{} observer.
 \item Finally, we notice that {the} BF magnitude increase observed {in GR} for {a} sparse sample of \samIII{} is stronger 
 {than the} non-GR effect of our MG model.
 \end{enumerate}

\end{itemize}

\section{Discussion}
\label{sec:discussion}

The above summary of all our important findings regarding various systematic effects that impact low-order moments of the galaxy velocity field,
underlines a number of crucial observations. The linear theory predictions (for MLV) obtained {using} 
Eqns.(\ref{eqn:BF_linear_theory}-\ref{eqn:BF_distribution})
{render quite} accurate values of BF {for a ``cosmic mean''} (\ie{} {\it Copernican}) observer in the case of high-density clean data.
Thus, they can be used as a first order prediction for the case when all {systematic} effects can be ignored, or when there are no computer
{simulations} to be compared with. {However, we caution that} whenever one wants to compare such LT predictions 
to the real data{, one needs to remember that the local galaxy velocity field is biased w.r.t. the LT prediction at small scales
and for sparse or radially incomplete samples.} 

Currently,
the velocity data is sparse and noisy;
however, in the near future they will 
increase significantly in volume. There is also hope for better modeling and understanding
of galaxy intrinsic scaling relations, which can lead to further suppression of individual velocity errors.
The surveys such as TAIPAN \cite{Taipan} or WALLABY \cite{WALLABY} 
{will lead to an increase both in} richness as well {as depth of galaxy peculiar velocity catalogs}. In addition, 
{the possibility to obtain} 
transverse velocities from surveys like GAIA {\citep{Gaia2012}} or LSST {\citep{LSST}}
{ could result in additional largely} unbiased {velocity measurements for the nearby galaxies 
(for a dedicated discussion see \cite{Nusser2012}). 
In this new era of velocity data,} the linear theory predictions for \BF{} and related statistics will be too inaccurate 
to be used for model testing and data analysis, {except for the large scales (\ie{} $\geq100-150\hmpc$), where the precision
and data abundance will continue to be poor}.

The various observer-independent systematic effects surfacing strongly {in} our analysis suggest that {the} bulk flow amplitude and related measurements 
at small distances
should be carefully reanalyzed and compared with predictions based on galaxy-mock catalogs. 
The higher BF values 
reported by Refs.~\citep{Branchini2012,Hong2014,Kashlinsky2008,Ma2013,Ma2014,Hoffman2015} might be a signature of biases induced by
sparse sampling and radial selection. \

The importance of proper modeling of such non-linear effects is {of paramount} importance for the cosmic Mach number predictions. This was
already emphasized by Ref.~\cite{Agarwal2013} for the case of improving the {LT} predictions by using {the} non-linear matter 
{density} power spectrum rather
{than the} linear one.  Here, our analysis {adds} further that sparse sampling induces a 
very strong effect on {the} VD and thus on the resulting CMN.
In addition, other {systematics such} as radial selection {and velocity} errors affect $\MN$~ to {a} much stronger extent than BF.
{This suggests in particular} that one should aim {at} using the richest {possible} galaxy {samples when} considering CMN measurements.

The {proximity} of a Virgo-like {cluster to a Local-Group-like} observer is {equally} significant and needs to be considered as 
an additional important
contribution to the local bulk flow. If this effect is not properly accounted for in the {BF analysis,} it will {result in} an additional 
{non-Gaussian systematic for} the BF measured {on} scales $R\simlt100\hmpc$.

Finally, the combination of  all the aforementioned systematic effects, if not accounted for carefully, can lead to strong degeneracies
of the cosmological signals encoded in galaxy velocities and {in} their low-order moments. We have clearly demonstrated that the signal of
{a} non-GR cosmological model, such as the nDGP modified gravity we considered, that employs a moderately strong modification to the cosmic
growth rate of structures, can be easily absorbed by the non-trivial systematics effects we {studied}. In light of this {evidence, analyses such as} 
for example Ref. \cite{SeilerParkinson2016}, where the BF deviations induced by modified gravity {were} studied,
should be definitely revisited.

All this is a source of potentially major concern,  as the cosmic velocity data {offers, at least} in principle, 
a model-independent way to constrain growth rate and gravity on cosmic scales
{\citep{Hellwing2014PhRvL,Feix2014,Hoffman2015,Nusser2016,Hoffman2016}}. {Other authors have also shown that current and future} data 
show promise to become
competitive cosmological probes {\citep{Koda2014,Howlett2017}}. In principle, {this can} still be achieved. However, the results
presented here clearly indicate that all the {various systematic effects} need to be carefully {addressed} and accounted for, before any
high-accuracy cosmological analysis can be {performed. This is especially important that in the near future, the amount of peculiar 
velocity data is expected to significantly increase, and therefore systematics will likely dominate over statistical errors 
in relevant studies.} In this context, {using} dedicated computer simulations employing constrained local
density (and velocity) realizations 
(such as \citep{Gottloeber2010,Courtois2012,Hess2013,Sorce2014,Leclercq2015,Sawala2016,Sorce2016,Leclercq2017,Desmond2018}) look very promising.

\section*{Acknowledgments}
The authors are very grateful to Baojiu Li for inspiring discussions and for providing the {\tt ECOSMOG} code 
that was used to run the simulations in this paper. 
Enzo Branchini, Martin Feix and Adi Nusser are also acknowledged for stimulating discussions and for careful reading
of the manuscript.
WAH is supported by an Individual Fellowship of the Marie Skłodowska-Curie Actions and therefore acknowledges that
this project has received funding from the European Union’s Horizon 2020 research and innovation
program under the Marie Skłodowska-Curie grant agreement No 748525.
MB is supported by the Netherlands Organization for Scientific Research, NWO, through grant number 614.001.451, 
and by the Polish National Science Center under contract UMO-2012/07/D/ST9/02785.
This project has also benefited from numerical computations performed at 
the Interdisciplinary Centre for Mathematical and Computational Modelling (ICM) University of Warsaw 
under grants no GA67-17 and GA65-30.
This work used the COSMA Data Centric system at Durham University, operated by the Institute for 
Computational Cosmology on behalf of the STFC DiRAC HPC Facility (www.dirac.ac.uk. This equipment 
was funded by a BIS National E-infrastructure capital grant ST/K00042X/1, DiRAC Operations grant ST/K003267/1 
and Durham University. DiRAC is part of the National E-Infrastructure.

\section*{Appendix}
\appendix
\label{sec:appendix}
\begin{figure*}
 \includegraphics[angle=0,width=0.96\textwidth]{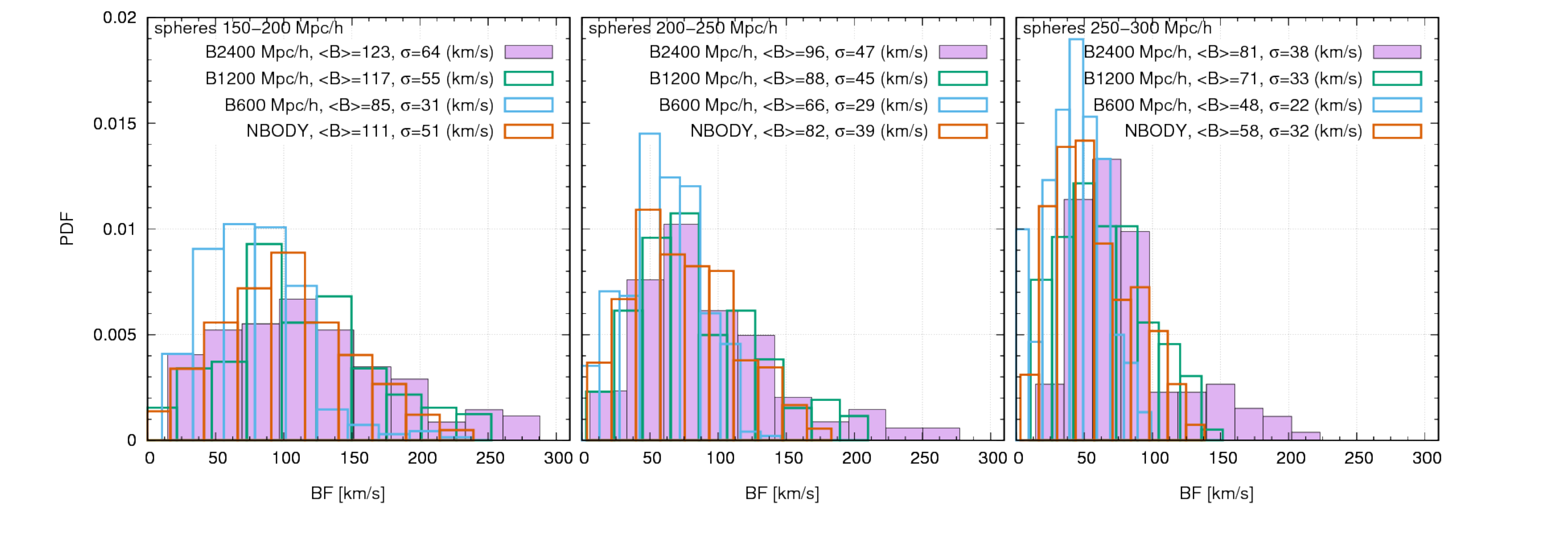}
 \caption{The distribution of BF magnitudes for three different spheres for our COLA runs. The left-hand side panel considers
 spheres of radii from $150-200\hmpc$, the middle panel is for a range of $200-250\hmpc$, while the right-hand side
 plot illustrates the results for $250-300\hmpc$. In each panel the filled boxes are for the fiducial $2400\hmpc$ box, the open green
 boxes mark the $1200\hmpc$ run and the open blue boxes the $600\hmpc$ one. In each label a median value $\langle\BF\rangle$ 
 for a given distribution is given together with $\sigma$, which marks a corresponding distribution spread between the $16$-th 
 and $84$-th percentiles.}
\label{fig:BF_PDF_BOXES}
\end{figure*}
{
Here we will investigate how and on what scales the limited simulation volume affects our measurements. As mentioned in
the main text, the cosmic velocity field is characterized by a large correlation length. This means that the convergence
of velocity moments is slower than in the case of the density field. For that reason we can expect that scales which are normally
considered as converged will be still affected by missing large-scale modes in our simulations. To assess this we have conducted a
series of auxiliary approximated simulations varying the box size. For this we employ 
the {\it COmoving Lagrangian Accelerator} (COLA) method \citep{Tassev2013}. 
The parallel implementation of the COLA algorithm (called PI-COLA, see \citep{Howlett2015}) allows to run large simulations 
at a reduced computational cost, with the trade-off of limited spatial and temporal resolution. 
We are however here interested in the effect of the missing large-scale modes, thus the scales which we will study, namely $100-300\hmpc$,
are large enough to be fully resolved by the PI-COLA method. In particular, we have used the publicly available optimized branch of the COLA
family, the MG-COLA, introduced by Ref.~\citep{Winther2017}.}
\begin{figure*}
 \includegraphics[angle=0,width=0.96\textwidth]{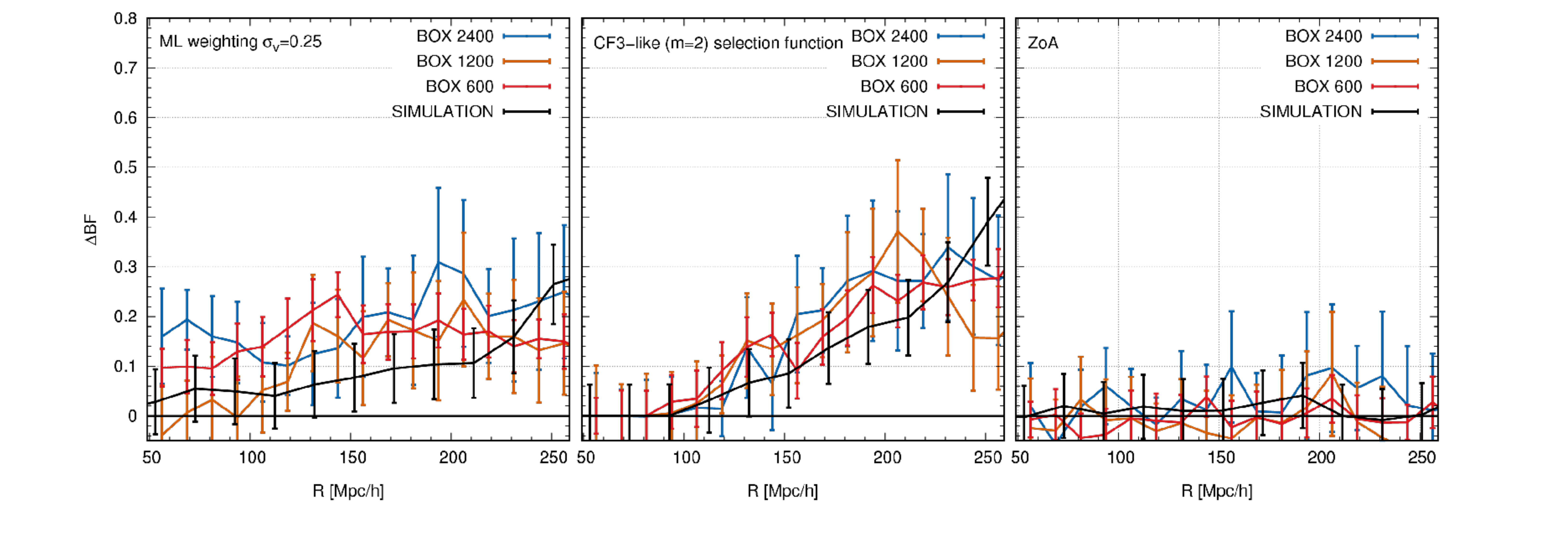}
 \caption{The relative differences $\Delta$BF taken between the fiducial full sample $\BF(R)$ and for various data weighting effects. Each panel
 compares the results for three COLA boxes and our $N$-body simulation, while the particular modeled effect varies from panel to panel.
 The left-hand side considers ML individual velocity error weighting, the middle plot shows a CF3-like radial selection with $m=2$, and  the right-hand side panel presents the effect of ZoA modeling. In each panel the error bars mark the error on the relative ratio computed 
 using the bootstrapped errors on the median \BF ~for 100 random observers.}
\label{fig:BF_weights_BOXES}
\end{figure*}

{We ran three simulations, all containing $1600^3$ volume elements, with three boxes: $600, 1200$ and $2400\hmpc$ on a side. The simulations
are set to use the same cosmology as our main N-body runs used in this work. We process the simulation outputs at $z=0$ in the same manner
as our full N-body simulations. We use the final ROCKSTAR halo catalogs as our input data. The COLA method is known
to bias weakly the resulting halo velocities. However, this bias is small (up to $\sim5\%$ for our case) and concerns mostly the small-scale
halo velocity field (see more in Ref.~\citep{Munari2017}). Since our primary concern here is to study the effects of missing large-scale power,
we are confident that halo catalogs obtained via the simplified COLA method are suitable for our purpose.}

{In FIG.~\ref{fig:BF_PDF_BOXES} we show probability distribution functions of \BF\ magnitudes for our COLA runs computed and binned for spheres
of three radii: $150-200\hmpc$ in the left-hand panel, $200-250\hmpc$ for the middle one, and $250-300\hmpc$ for the right-hand panel. The PDF 
for the fiducial run of $2400\hmpc$ box is illustrated by filled purple boxes, while two consecutively smaller simulations are depicted
by open  green ($1200\hmpc$) and blue ($600\hmpc$) boxes. For comparison we also plot the relevant PDFs for the full $N$-body simulation used
in the paper. The immediate impression is that all the results for the smallest box are significantly
biased w.r.t. the fiducial case. Both the median of the distribution ($\langle\BF\rangle$), as well as its spread ($\sigma$)
\footnote{For our purpose here, for a measure of the distribution spread ({\it i.e.}~$\sigma$) we 
use the half-width between the 16-th and 84-th percentiles.} are visibly smaller for all three radii.
The corresponding relative differences taken w.r.t the fiducial case here (\ie~the largest box) 
are $\sim30\%$ ($31\%$) and $\sim50\%$ ($38\%$) for the median and spread 
of the PDF at $150-200\hmpc$ ($200-250\hmpc$) and $\sim40\%$ and $\sim42\%$ at $250-300\hmpc$. 
This is a clear manifestation of the lacking large-scale
power in the simulation box. Thus we can, not surprisingly, conclude that the halo velocity field is not converged at those scales 
in the $600\hmpc$ box. The situation for the medium box $1200\hmpc$ is much better though. Here, the medians are off by only
$\sim6\%$ ($\sim8\%$) at $150-200\hmpc$ ($200-250\hmpc$), while the corresponding distribution widths are smaller by $\sim14\%$ ($\sim5\%$).
However, at the largest radius of $250-300\hmpc$ the biases grow to $\sim12\%$ for the median and $\sim13\%$ for the width.
Our full $N$-body simulations use a $1000\hmpc$ box, and it is reassuring to find biases of their adequate $\BF$ distributions to lie
between COLA $1200$ and $600\hmpc$ boxes. At $200\leq R(\hmpc)^{-1}\leq 250$ the median is biased by more than $14\%$ already.
This shows that all the results in this paper for spheres larger than $R\simgt 200\hmpc$ are noticeably affected by missing
large-scale power.}

{Obtaining a reliable and accurate absolute bulk flow (and corresponding dispersion) magnitude is of paramount importance when one
wants to compare it with astronomical data and use such a comparison for parameter constraints. In this paper however, we are more 
interested in some specific effects that affect the velocity field statistics in a systematic way. Thus, to study to what scales
we can trust our results we present FIG.~\ref{fig:BF_weights_BOXES}. Here, we compare size of relative differences
(taken always w.r.t. the fiducial unweighted sample) for three systematic effects: ML velocity error weights (left panel),
CF-3-like (with $m=2$) radial selection function (middle panel) and the effect of the Zone of Avoidance (right panel).
Reassuringly, we denote that in all three cases the scale-dependence as well as $\Delta\BF$~magnitude are very similar (within the sampling error)
for the three COLA run and our full $N$-body. The largest differences appear for the individual velocity error weights. Here,
the $N$-body results for $R\simgt200\hmpc$ are consistently $1\sigma$ below the $2400\hmpc$ COLA line. This indicates that for the case of this
systematic effect and at those scales our results render only the lower-bound, and amore realistic modeling will probably foster
larger effects.}

{Finally, we can report that the $\BF(R)$~ distributions (both from COLA and $N$-body) at all probed scales are deviating significantly
from a Gaussian, with  typical skewness ($S_3$) and kurtosis ($S_4$) taking values $|S_3|,|S_4|\sim \mathcal{O}(1)$.}

\renewcommand{\bibname}{References}
\bibliography{uneven_flows}
\end{document}